\documentclass[12pt]{article}
\usepackage{amssymb}
\usepackage{amsmath}
\usepackage{hyperref}
\usepackage{graphicx}
\usepackage[font=small,labelfont=bf]{caption}
\begin{document}
\title{\bf  The Phase Transition of Non-minimal Yang-Mills AdS Black Brane}
\author{ 
      Mehdi Sadeghi\thanks{Corresponding author: Email:  mehdi.sadeghi@abru.ac.ir} and Faramaz Rahmani\thanks{Email:  faramarz.rahmani@abru.ac.ir}\hspace{2mm}\\
{\small {\em Department of Physics, School of Sciences,}}\\
        {\small {\em Ayatollah Boroujerdi University, Boroujerd, Iran}}
       }
\date{\today}
\maketitle

\abstract{In this paper, we shall study the phase transition of non-minimal coupling of Einstein-Hilbert gravity and electric field of Yang-Mills type in AdS space-time. We couple the Ricci scalar to the Yang-Mills invariant to obtain a modified theory of gravity. A black brane solution is introduced up to the first order of the term $RF^{(a)}_{\mu \alpha }F^{(a)\mu \alpha} $ in this model. Then, the phase transition of this solution will be investigated in canonical ensemble. Our investigation shows that only the second order phase transition behavior is seen in this model. Also, due to the coupling of the Yang-Mills field and Ricci scalar, there are differences with the phase transitions of the usual minimal models. We shall show that in the absence of non-minimal coupling there is no any phase transition.}\\



\section{Introduction} \label{intro}
General relativity (GR)is a successful theory because it passed the gravitational tests well. The black holes which are singular solutions of Einstein's equations with event horizons are the most interesting prediction of GR. The existence of black holes was confirmed by Laser Interferometer Gravitational Wave-Observatory (LIGO) and the Virgo detectors \cite{LIGOScientific:2021qlt}. The event horizon telescope collaboration published the first image of a black hole in 2019.\par 
The non-minimal theories which couple the gravitational field to other fields, were introduced as  alternative theories of gravity\cite{Sadeghi:2021qou},\cite{Sadeghi:2022mog},\cite{Balakin:2005fu},\cite{Balakin:2015gpq}. In this paper, we study the first class of non-minimal theory which couples the gravitational field to scalar fields\cite{Balakin:2015gpq}. Power-law inflation can be understand by the non-minimal gravitational coupling of the electromagnetic field\cite{Bamba:2008ja}. Therefore, dark matter and dark energy as unknown parts of the universe can be realized by non-minimal models.\par 
A black hole behaves as a thermodynamic system, because it has entropy, temperature, pressure and in some situations it experiences phase transition.\cite{Hawking:1982dh}-\cite{Ranjbari:2019ktp}. The temperature of a black hole, $T=\frac{\mathcal{K}}{2 \pi}$, is called Hawking temperate where $\mathcal{K}$ denotes the surface gravity of the black hole\cite{Hawking:1982dh}. Hawking investigated the black hole radiation through the study of quantum field theory in curved space-time. In the extended phase space approach, the pressure of a black hole  is related to the cosmological constant through the relation, $P=-\frac{\Lambda}{8\pi G}$ and in $d$-dimensions we have $\Lambda=\frac{(d-1)(d-2)}{2L^2}$. Also, the mass of the black hole plays the role of enthalpy. The behavior of Gibbs free energy classifies the phase transitions of the black holes \cite{Hendi:2017mfu}. The van der Waals-like phase transition \cite{Chamblin:1999tk},\cite{Chamblin:1999hg},\cite{Wei:2010yw} can be seen in charged black hole systems by considering the correspondence ($Q$,$\Phi$)$\leftrightarrow$ ($P$, $V$) between conserved quantities and thermodynamic variables. Thermodynamic volume, $V$, is the conjugate quantity to the pressure through the relation $V=(\frac{\partial M}{\partial P})$.\par 
AdS/CFT duality \cite{Maldacena},\cite{Aharony},\cite{Witten:1998qj} opens a new window to study the strongly coupled systems. Gravity is located in the bulk of AdS, while field theory is located on the boundary of AdS. These two different theories are related through the dictionary of this duality. According to dictionary of AdS/CFT duality, the first order phase transition in the bulk corresponds to the confinement-deconfinement transition \cite{Witten:1998zw}  on the boundary and the second order phase transition in the bulk corresponds to a phase transition from a normal phase to a superconducting phase on the boundary \cite{Gubser:2005ih}-\cite{Gubser:2008px}.\par 
In this paper, we want to study the non-minimal Yang-Mills black brane with a cosmological constant and study the holographic aspects through the investigation of phase transition of the system. In the next section, we shall obtain a perturbed solution for the non-minimal $RF^{(a)}_{\mu \alpha }F^{(a)\mu \alpha}$ AdS black brane. Then, the thermodynamics of the solution will be studied in the section (\ref{sec3}).
\section{  Non-minimal $RF^{(a)}_{\mu \alpha }F^{(a)\mu \alpha} $  AdS Black Brane Solution}
\label{sec2}
The action of the regular non-minimal $RF^{(a)}_{\mu \alpha }F^{(a)\mu \alpha} $ Yang-Mills theory in four dimensions is written in the form\cite{Balakin:2005fu},\cite{Shepherd:2015dse}:
\begin{eqnarray}\label{action}
	S=\int d^{4}  x\sqrt{-g} \bigg[\frac{1}{\kappa }(R-2\Lambda )+\frac{q_1}{2}F^{(a)}_{\mu \alpha }F^{)(a)\mu \alpha} +q_2 RF^{(a)}_{\mu \alpha }F^{(a)\mu \alpha} \bigg],
\end{eqnarray}
where $R$ is the Ricci scalar, $\Lambda=-\frac{3}{L^2}$ is the cosmological constant, $L$ is the AdS radius, $\mathcal{F}={\bf{Tr}}( F_{\mu \nu }^{(a)} F^{(a)\, \, \mu \nu })$  is the Yang-Mills invariant and $\kappa=8 \pi G$. Here, $ F^{(a)\, \, \mu \nu }$ is the Yang-Mills field tensor,
\begin{align} \label{YM}
	F_{\mu \nu } =\partial _{\mu } A_{\nu } -\partial _{\nu } A_{\mu } -i[A_{\mu }, A_{\nu }],
\end{align}
where, the gauge coupling constant is taken equal to unity and $A_{\nu }$'s are the Yang-Mills potentials. Here, $q_2$ is the dimensionless coupling for the interaction between the gauge field and the Ricci scalar.
The equations of motion are obtained through the variation of  action (\ref{action}) with respect to the metric. The equations are as follows:
\begin{equation}\label{EOM1}
	R_{\mu \nu }-  \tfrac{1}{2} g_{\mu \nu } R + \Lambda g_{\mu \nu }=\kappa T^{\text{(eff)}}_{\mu \nu },
\end{equation}
where,
\begin{equation}
	T^{\text{(eff)}}_{\mu \nu }=q_1T^{\text{(YM)}}_{\mu \nu } + q_2T^{(I)}_{\mu \nu },
\end{equation}
\begin{equation}
	T^{\text{(YM)}}_{\mu \nu }=  \tfrac{1}{4}g_{\mu \nu } F^{(a)}_{\alpha \beta } F^{^{(a)} \alpha \beta }-F_{\mu }{}^{(a)\alpha } F^{(a)}_{\nu \alpha },
\end{equation}
\begin{eqnarray}
	&&T^{(I)}_{\mu \nu }=\tfrac{1}{2} F^{(a)}_{\alpha \beta } F^{(a)\, \alpha \beta} g_{\mu \nu  } R-R_{\mu \nu } F^{(a)}_{\alpha \beta } F^{^{(a)} \alpha \beta } - 2 F_{\mu }^{(a) \,\alpha} F^{(a)}_{\nu \alpha }  R \nonumber \\ 
	&&-2 F^{(a)}_{\alpha \beta }  g_{\mu \nu  }\nabla_{\gamma }\nabla^{\gamma }F^{(a)}_{\alpha \beta }-2 g_{\mu \nu  } \nabla_{\gamma }F^{(a)} _{\alpha \beta } \nabla^{\gamma }F^{(a) \,\alpha \beta } + F^{(a) \,\alpha \beta } \nabla_{\mu } \nabla_{\nu } F^{(a)}_{\alpha \beta } \nonumber \\ 
	&& +2 \nabla_{\mu } F^{(a) \,\alpha \beta } \nabla_{\nu } F^{(a)}_ {\,\alpha \beta }+F^{(a) \,\alpha \beta } \nabla_{\nu } \nabla_{\mu } F^{(a)}_{\alpha \beta }.
\end{eqnarray}
The equations of the Yang-Mills field are given through the variation of the action (\ref{action}) with respect to the  $A_{\mu}$ as follows:
\begin{eqnarray}\label{EOM-Maxwell}
	\nabla_{\mu }\Big(q_1 F^{(a)\mu \nu } + 2 q_2 F^{(a)\mu \nu } R\Big)=0.
\end{eqnarray}
We want to find the black brane solution of this model. So, we suggest the following ansatz:
\begin{equation}\label{metric}
	ds^{2} =-e^{-2H(r)}f(r)dt^{2} +\frac{dr^{2} }{f(r)} +\frac{r^2}{L^2}(dx^2+dy^2).
\end{equation}
In order to solve the Eq.(\ref{EOM-Maxwell}), we choose the electric gauge potential in the form:
\begin{equation}\label{background}
	{\bf{A}}^{(a)} =\frac{i}{2}h(r)dt\begin{pmatrix}1 & 0 \\ 0 & -1\end{pmatrix}.
\end{equation}
The gauge group is the diagonal generator of the Cartan subalgebra of $SU(2)$ \cite{Shepherd:2015dse}.\par 
These equations do not have analytic solutions. Therefore, we solve them up to the first order of $q_2$ as introduced in \cite{Sadeghi:2023hxd}. We consider the following forms for $f (r)$, $H(r)$ and $h (r)$.
\begin{equation}\label{f}
	f(r)=f_0(r)+q_2 f_1(r),
\end{equation}
\begin{equation}\label{h}
	h(r)=h_0(r)+q_2 h_1(r),
\end{equation}
\begin{equation}\label{H}
	H(r)=H_0(r)+q_2 H_1(r),
\end{equation}
where, $f_0(r)$, $h_0(r)$ and $H_0(r)$ are the leading order solutions of the Einstein-Yang-Mills AdS black brane in four dimensions. We have
\begin{equation}
	h_0(r)=C_2-C_1\int^r\frac{ 1}{q_1 u^2}du=Q(\frac{1}{r}-\frac{1}{r_h}),
\end{equation}
\begin{equation}
	H_0(r)=0.
\end{equation}
Here, $Q=\frac{C_1}{q_1}$ and $C_2=\frac{Q}{r_h}$. In the following, we have:
\begin{equation}\label{f0}
	f_0(r)=-\frac{2m_0}{r}-\frac{\Lambda r^2}{3}+\frac{\mathit{k} q_1}{2 r}\int^r  u^2 
	h_0'^2 du=-\frac{2m_0}{r}+\frac{r^2}{L^2}-\frac{\mathit{k} q_1 Q^2}{2 r^2},
\end{equation}
where $m_0$ is
\begin{equation}\label{m}
	m_0=-\frac{\mathit{k} q_1 Q^2}{4 r_h}+\frac{r_h^3}{2L^2}.
\end{equation}
Also,  functions $f_1(r)$, $h_1(r)$ and $H_1(r)$ are given by,
\begin{equation}
	h_1(r)=-\mathit{k} q_1  Q^3(\frac{1}{r^5}-\frac{1}{r_h^5})-\frac{24 Q }{L^2  }(\frac{1}{r_h}-\frac{1}{r}),
\end{equation}
\begin{equation}\label{h1r}
	H_1(r)=C_3+ 2  \mathit{k} r h_0' h_0''- \mathit{k} h_0'^2=-\frac{5 \mathit{k} Q^2}{r^4},
\end{equation}
and
\begin{eqnarray}
	&&f_1(r)=-\frac{5 \kappa ^2 Q^4 q_1}{r^7} (1-\frac{r^6}{r_h^6})-\frac{\kappa ^2 Q^4 q_1}{2 r^6 r_h^5}(7 r r_h^4+r^5-8 r_h^5)\nonumber \\&&+\frac{\kappa  Q^2 }{L^2 r^5 r_h}\left(2 r^3 r_h-9 r^4+7 r_h^4\right)+\frac{24 \kappa  Q^2 \left(r^2-r_h^2\right)}{L^2 r^3 r_h^2}.
\end{eqnarray}
In relation (\ref{h1r}), $C_3=0$, which we have used the fact that the speed of light should be equal to unity on the boundary of AdS \cite{Sadeghi:2023hxd}.
Blackening factor on the event horizon vanishes. In other words, $f(r_h)=0$.
In the following, we investigate the thermodynamics of the system. 
\section{The thermodynamics of the model}
\label{sec3}
In order to study the thermodynamic properties of the system, we need to obtain the ADM mass of the system which plays the role of enthalpy in this context. The metric component $g_{tt}$ up to the first order of $q_2$ can be written as follows:
\begin{equation}
	g_{tt}=-e^{-2H(r)}f(r)\simeq -f_0-q_2f_1+2q_2f_0H_1.
\end{equation}
This leads to
\begin{eqnarray}\label{mm}
	&&g_{tt}=\frac{r^{2}}{L^{2}}+\frac{-\left(\frac{5 \kappa^{2} Q^{4} q_{1}}{r_{h}^{6}}-\frac{\kappa^{2} Q^{4} q_{1}}{2 r_{h}^{5}}-\frac{9 \kappa  Q^{2}}{L^{2} r_{h}}+\frac{24 \kappa  Q^{2}}{L^{2} r_{h}^{2}}\right) q_{2}-\frac{\kappa  q_{1} Q^{2}}{2 r_{h}}+\frac{r_{h}^{3}}{L^{2}}}{r}\nonumber \\&&+\frac{\frac{8 \kappa  Q^{2} q_{2}}{L^{2}}+\frac{\kappa  q_{1} Q^{2}}{2}}{r^{2}}+\frac{24 \kappa  Q^{2} q_{2}}{L^{2} r^{3}}+\frac{-10 \kappa  Q^{2} \left(\frac{\kappa  q_{1} Q^{2}}{2 r_{h}}-\frac{r_{h}^{3}}{L^{2}}\right) q_{2}-\left(-\frac{7 \kappa^{2} Q^{4} q_{1}}{2 r_{h}}+\frac{7 \kappa  Q^{2} r_{h}^{3}}{L^{2}}\right) q_{2}}{r^{5}}\nonumber \\&&+\frac{\kappa^{2} Q^{4} q_{1} q_{2}}{r^{6}}+\frac{5 \kappa^{2} Q^{4} q_{1} q_{2}}{r^{7}}.
\end{eqnarray}
We represent the numerator of the second term in relation (\ref{mm}) by $2m$ where, $m$ is an integration constant. 
In 4-dimensional models, ADM mass or enthalpy is equal to the integration constant $m$ \cite{Deh1}.
In other words, we have
\begin{equation}\label{ml}
	M=m=\left(\frac{9 \kappa  Q^{2}}{2 L^{2} r_{h}}-\frac{12 \kappa  Q^{2}}{L^{2} r_{h}^{2}}+\frac{\kappa^{2} Q^{4} q_{1}}{4 r_{h}^{5}}-\frac{5 \kappa^{2} Q^{4} q_{1}}{2 r_{h}^{6}}\right) q_{2}-\frac{\kappa  q_{1} Q^{2}}{4 r_{h}}+\frac{r_{h}^{3}}{2 L^{2}}=m_0+\delta m.
\end{equation}
This is the mass or enthalpy of the system up to the first order of perturbation. The mass of the system up to the zeroth order of perturbation, is the same as equation (\ref{m}). 
On the other hand, the cosmological constant, its associated pressure and the AdS radius are related through the relations,
\begin{equation}
	P=-\frac{\Lambda}{8 \pi G},
\end{equation}
and
\begin{equation}\label{l2}
	L^2=-\frac{3}{\Lambda}=\frac{3}{8\pi G P}.
\end{equation}
In terms of black brane pressure, relation (\ref{ml}) takes the following form ($G=1$ and $\kappa=8\pi$):
\begin{equation}\label{mp}
	M=\left(-\frac{160 \pi^{2} Q^{4} q_{1}}{r_{h}^{6}}+\frac{16 \pi^{2} Q^{4} q_{1}}{r_{h}^{5}}+\frac{96 Q^{2} \pi^{2} P}{r_{h}}-\frac{256 Q^{2} \pi^{2} P}{r_{h}^{2}}\right) q_{2}-\frac{2 \pi  q_{1} Q^{2}}{r_{h}}+\frac{4 r_{h}^{3} \pi  P}{3}.
\end{equation}
The entropy of the black brane can be calculated by the area law formula,
\begin{equation}\label{a1}
	S=\frac{A}{4G}=\frac{1}{4G} \int d^{2}x \sqrt{-g}\mid_{r=r_h,t=cte}.
\end{equation}
A direct calculation by using the metric components in relation (\ref{metric}), leads to
\begin{equation}\label{a2}
	S=\frac{r_h^2}{4}+\frac{10 \pi Q^2 q_2}{r_h^2}.
\end{equation}
\begin{figure}[h!]
	\centerline{\includegraphics[width=6cm]{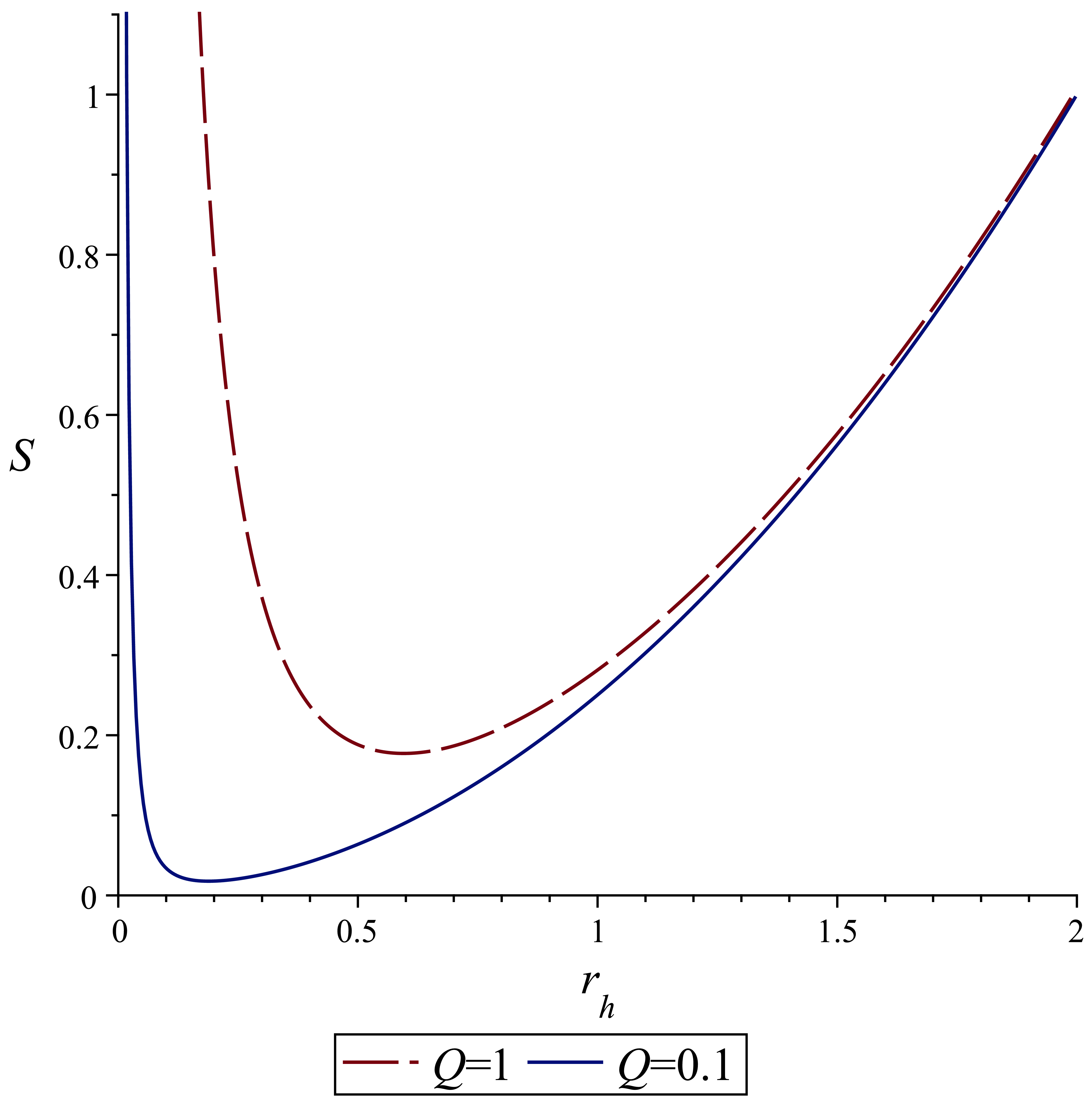}}
	\caption{The diagram of the perturbed entropy for the values $ Q=0.1, Q=1$ and $q_2=0.001$. \label{fig:scom}}
\end{figure}\\
This is a perturbative result for the entropy of the system. The appearance of $r_h$ in the denominator of the second term, makes some difficulties in the study of thermodynamics of the system. We focus only on the parameter values where $r_h$ is large. If the second term has a special physical meaning, we will not discuss it in this paper. We examine the problem in areas where the second term can be ignored. In other words, to have better results, the condition $r_h > (10 \pi q_2)^{\frac{1}{2}} Q$ should be satisfied minimally.
We shall see latter that a better estimate can be made by using the volume equation of the system. 
In Fig (\ref{fig:scom}) the diagram of the perturbed entropy for the Yang-Mills charges $Q=0.1$ and $Q=1$ with $q_2=0.001$, can be seen. We see that there is a minimum value in each diagram for the horizon of the black brane. For horizons smaller than the minimum horizon, the entropy of the system increases. This is contrary to the common understanding that the entropy of the system increases by increasing the area of the event horizon. Sometimes, it is said that the zero entropy is not achievable due to the uncertainty principle. Perhaps, such term is due to the quantum effects. Anyway, we consider
\begin{equation}\label{s2}
	S=\frac{r_h^2}{4},
\end{equation}
as the entropy of system. We can say this relation becomes more reliable by increasing the horizon $r_h$. 
Now, by using relations (\ref{s2}) and (\ref{mp}), the enthalpy of the system takes the form:
\begin{equation}\label{ms}
	M=\left(-\frac{5 \pi^{2} Q^{4} q_{1}}{2 S^{3}}+\frac{\pi^{2} Q^{4} q_{1}}{2 S^{\frac{5}{2}}}+\frac{48 Q^{2} \pi^{2} P}{\sqrt{S}}-\frac{64 Q^{2} \pi^{2} P}{S}\right) q_{2}-\frac{\pi  q_{1} Q^{2}}{\sqrt{S}}+\frac{32 S^{\frac{3}{2}} \pi  P}{3}.
\end{equation}
The effect of $q_2$ on the mass or enthalpy of the system is seen in this relation. 
The Hawking temperature can be obtained by using the relation
\begin{equation}\label{haw}
	T=\frac{1}{2 \pi} \Big[ \frac{1}{\sqrt{g_{rr}}}\frac{d}{dr}\sqrt{-g_{tt}}\Big]\Bigg|_{r=r_h}=\frac{e^{-H(r_h)} f'(r_h)}{4 \pi}.
\end{equation}
The first law of thermodynamics in the extended phase space for this model becomes
\begin{equation}
	dM=TdS+VdP+\phi dQ,
\end{equation}
where,
\begin{equation}\label{tm}
	\begin{split}
		&T=\left(\frac{\partial M}{\partial S}\right)_{P,Q}=\\ &\left(\frac{15 \pi^{2} Q^{4} q_{1}}{2 S^{4}}-\frac{5 \pi^{2} Q^{4} q_{1}}{4 S^{\frac{7}{2}}}-\frac{24 Q^{2} \pi^{2} P}{S^{\frac{3}{2}}}+\frac{64 Q^{2} \pi^{2} P}{S^{2}}\right) q_{2}+\frac{\pi  q_{1} Q^{2}}{2 S^{\frac{3}{2}}}+16 \sqrt{S}\, \pi  P ,
	\end{split}
\end{equation}
\begin{equation}
	V=\left(\frac{\partial M}{\partial P}\right)_{S,Q}=\left(\frac{48 Q^{2} \pi^{2}}{\sqrt{S}}-\frac{64 Q^{2} \pi^{2}}{S}\right) q_{2}+\frac{32 S^{\frac{3}{2}} \pi}{3},
\end{equation}
and
\begin{equation}
	\begin{split}
		&\phi =\left(\frac{\partial M}{\partial Q}\right)_{S,P}=\\
		&\left(-\frac{10 \pi^{2} Q^{3} q_{1}}{S^{3}}+\frac{2 \pi^{2} Q^{3} q_{1}}{S^{\frac{5}{2}}}+\frac{96 Q \pi^{2} P}{\sqrt{S}}-\frac{128 Q \pi^{2} P}{S}\right) q_{2}-\frac{2 \pi  q_{1} Q}{\sqrt{S}}.
	\end{split}
\end{equation}
These relations in terms of black brane horizon are as follows:
\begin{equation}\label{tm2}
	\begin{split}
		&T=\left(-\frac{192 \pi^{2} P \,Q^{2}}{r_{h}^{3}}+\frac{1024 \pi^{2} P \,Q^{2}}{r_{h}^{4}}-\frac{160 \pi^{2} Q^{4} q_{1}}{r_{h}^{7}}+\frac{1920 \pi^{2} Q^{4} q_{1}}{r_{h}^{8}}\right) q_{2}+\\
		&8 \pi  r_{h} P +\frac{4 \pi  q_{1} Q^{2}}{r_{h}^{3}},
	\end{split}
\end{equation}
\begin{equation}\label{vr}
	V=\left(\frac{96 \pi^{2} Q^{2}}{r_{h}}-\frac{256 \pi^{2} Q^{2}}{r_{h}^{2}}\right) q_{2}+\frac{4 r_{h}^{3} \pi}{3},
\end{equation}
and
\begin{equation}
	\phi=\left(\frac{192 Q \pi^{2} P}{r_{h}}-\frac{512 Q \pi^{2} P}{r_{h}^{2}}+\frac{64 Q^{3} \pi^{2} q_{1}}{r_{h}^{5}}-\frac{640 Q^{3} \pi^{2} q_{1}}{r_{h}^{6}}\right) q_{2}-\frac{4 Q \pi  q_{1}}{r_{h}}.
\end{equation}
\par 
The Smarr relation in this model takes the form:
\begin{equation}
	M=2TS-2VP+Q\phi.
\end{equation}
The equation of state is derived through the relation (\ref{tm2}). The result is as follows:
\begin{equation}\label{p11}
	P = -\frac{160 \pi^{2} Q^{4} q_{1} q_{2} r_{h}-4 \pi  Q^{2} q_{1} r_{h}^{5}+T r_{h}^{8}-1920 \pi^{2} Q^{4} q_{1} q_{2}}{8 \pi  r_{h}^{4} \left(24 \pi  Q^{2} q_{2} r_{h}-r_{h}^{5}-128 \pi  Q^{2} q_{2}\right)}.
\end{equation}
In the following, we discuss about the critical behavior of the system.
\begin{figure}[h!]
	\centering
	[a]{\includegraphics[width=6cm]{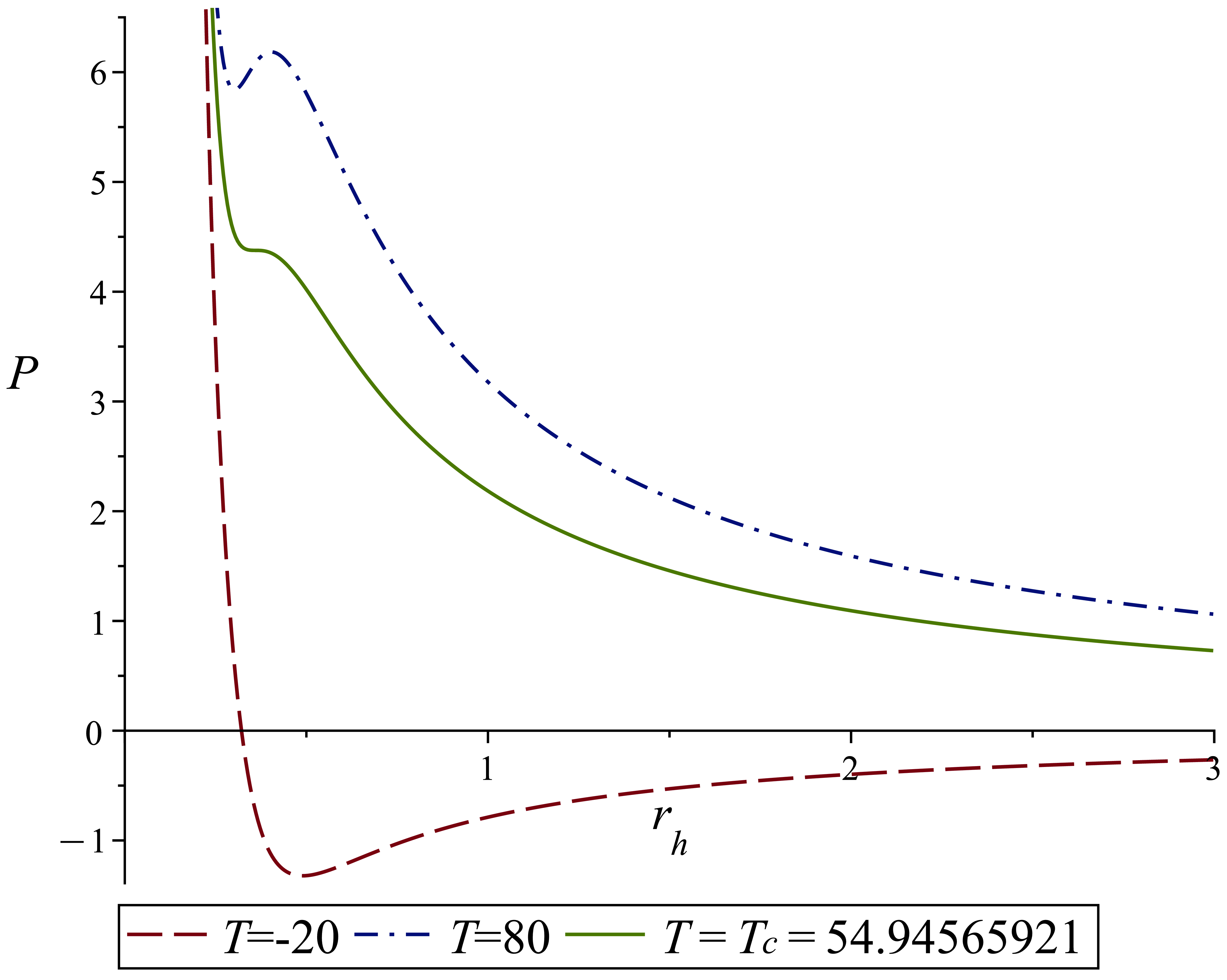}}\qquad
	[b]{\includegraphics[width=6cm]{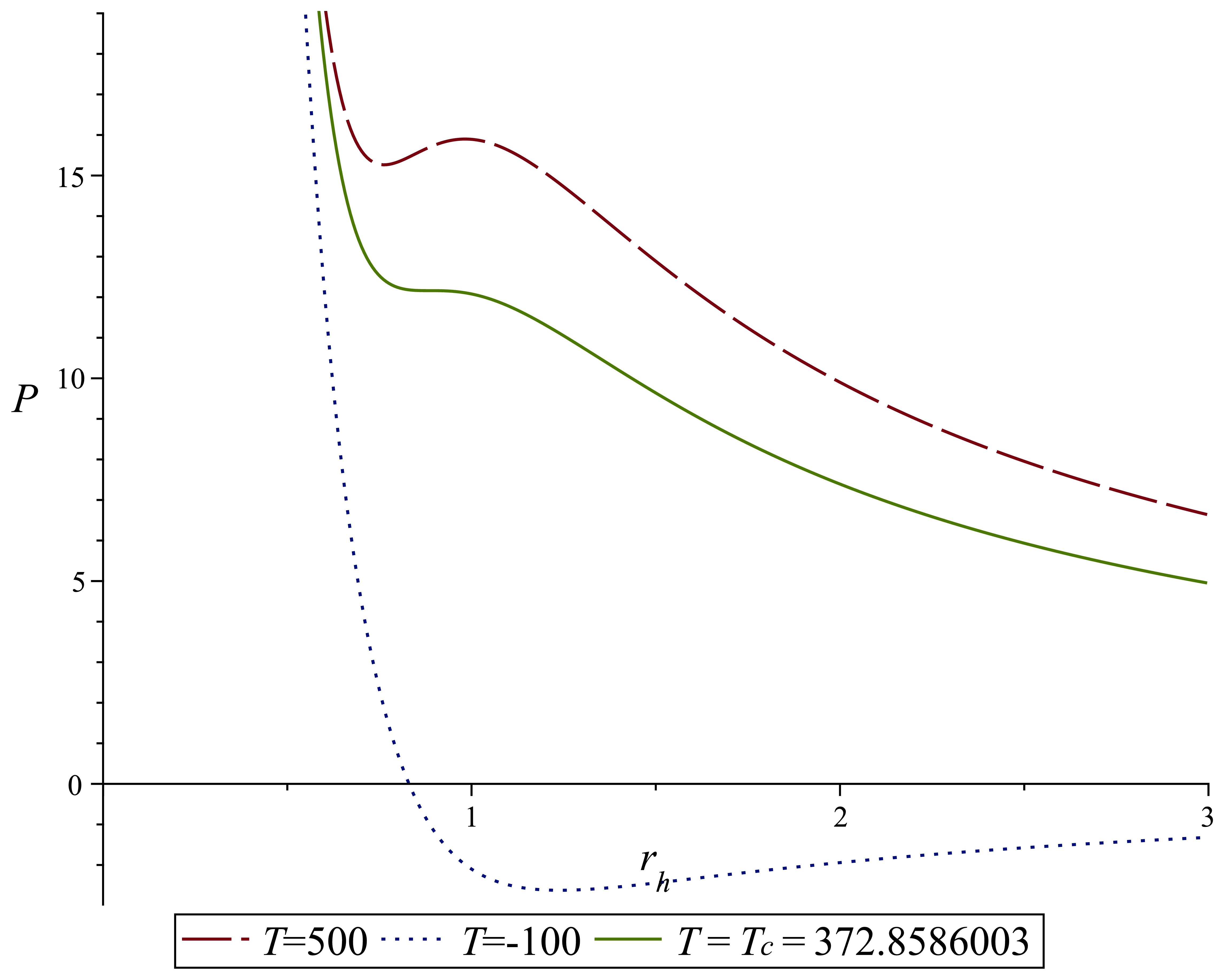}}
	\caption{$P-r_{h}$ diagram of the system; (a): for $ Q=0.1,q_1=-1$ and $q_2=0.001$. The solid line represents the critical behavior of the system for the critical values $r_c=0.3598661919$ and $T_c=54.94565912$., (b): for $Q=1,q_1=-1$ and $q_2=0.001$. The solid line represents the critical behavior of the system for the critical values $r_c=0.8868022599$ and $T_c=372.8586003$. \label{fig:2}}
\end{figure}
Before that, let's to have an estimation for the area we are talking about. By examining the volume formula (\ref{vr}), we realize that for some parameter values, the volume becomes negative. Unfortunately, a general analytical relation was not obtained. We realized that for values $Q=0.1, Q=1$ and $Q=2$, the values of $r_h$ for which the volume is positive are $0.3498444712$, $0.8381402930$ and $1.075589864$ respectively. Fortunately, these values are less than the value of the critical horizons that we will obtained latter. Thus, with charges $Q=0.1, Q=1$ and $Q=2$, this model is reliable for the horizons larger than the values $0.3498444712$, $0.8381402930$ and $1.075589864$ respectively. We mention that throughout the paper we have considered $q_2=0.001$ and $q_1=-1$.\par 
We know that at the critical point of the system, the condition
\begin{equation}\label{cr1}
	\frac{\partial P}{\partial r_h}|_{T=T_c,r_h=r_c}=0, \qquad  \frac{\partial^2 P}{\partial^2 r_h}|_{T=T_c,r_h=r_c}=0,
\end{equation}
should be satisfied. 
Unfortunately, the investigation of the problem shows that it is impossible to find some analytical relations between the critical values of the thermodynamic quantities in terms of the Yang-Mills charge as the only system parameter. Nevertheless, for some fixed Yang-Mills charge (canonical ensemble) we examine the critical behavior of the system. Our study shows that the critical behavior or phase transition can be seen in this system. For example, for the case $(Q=0.1,q_1=-1,q_2=0.001)$, the condition (\ref{cr1}) gives the critical values $r_c=0.3598661919$ and $T_c=54.94565912$, for the horizon and temperature respectively. These values give the critical value $P_c=4.356738199$ for the pressure of the system. The solid lines in Fig (\ref{fig:2}) show the isotherms of the system for critical temperatures. In the left panel, at $r_c=0.3598661919$ an inflection point is seen. For the Yang-Mills charge $Q=1$, an inflection point is seen in the right panel at $r_c=0.8868022599$.  
We expect that by increasing the temperature, the behavior of the system to be close to that of an ideal gas. This is true for large horizons. But, for small horizons, unstable behaviors are seen in the associated curves. For example, in the left panel the dash-dot curve with temperature $T=80>T_c$ shows unstable behavior for small horizons. This is not the behavior of an ideal gas for which the slope of $P-V$ curve is negative and the system is stable. This is due to the  coupling of the Yang-Mills field and the Ricci scalar  and can be interesting in its own way. Fortunately, the critical horizon in any case is larger than the minimum horizon for which the system volume is positive. In other words, the inflection points are in physical regions.
For different values of the Yang-Mills charge, the corresponding critical values can be found in the following table. 
\begin{table}[!h]
	\centering
	\caption{Critical quantities for the different Yang-Mills charge $Q$. Other parameters are $q_1=-1$ and $ q_2=0.001$.}
	\begin{tabular}{|c|c|c|c|}
		\hline
		$Q$ & $r_c$ & $P_c$ &  $T_c$  \\
		\hline  
		0.1 & 0.3598661919 & 4.376538199 &  54.94565912  \\
		\hline 
		1 &0.8868022599 & 12.16191079 &   372.8586003 \\
		\hline 
		2 &1.157739166 & 16.97567521 &   675.7415973 \\ \hline
	\end{tabular} 
\end{table}
This table shows that by increasing the Yang-Mills charge, the value of the critical horizons, critical pressures and critical temperatures increase.\par
\begin{figure}[h!]
	\centerline{\includegraphics[width=6cm]{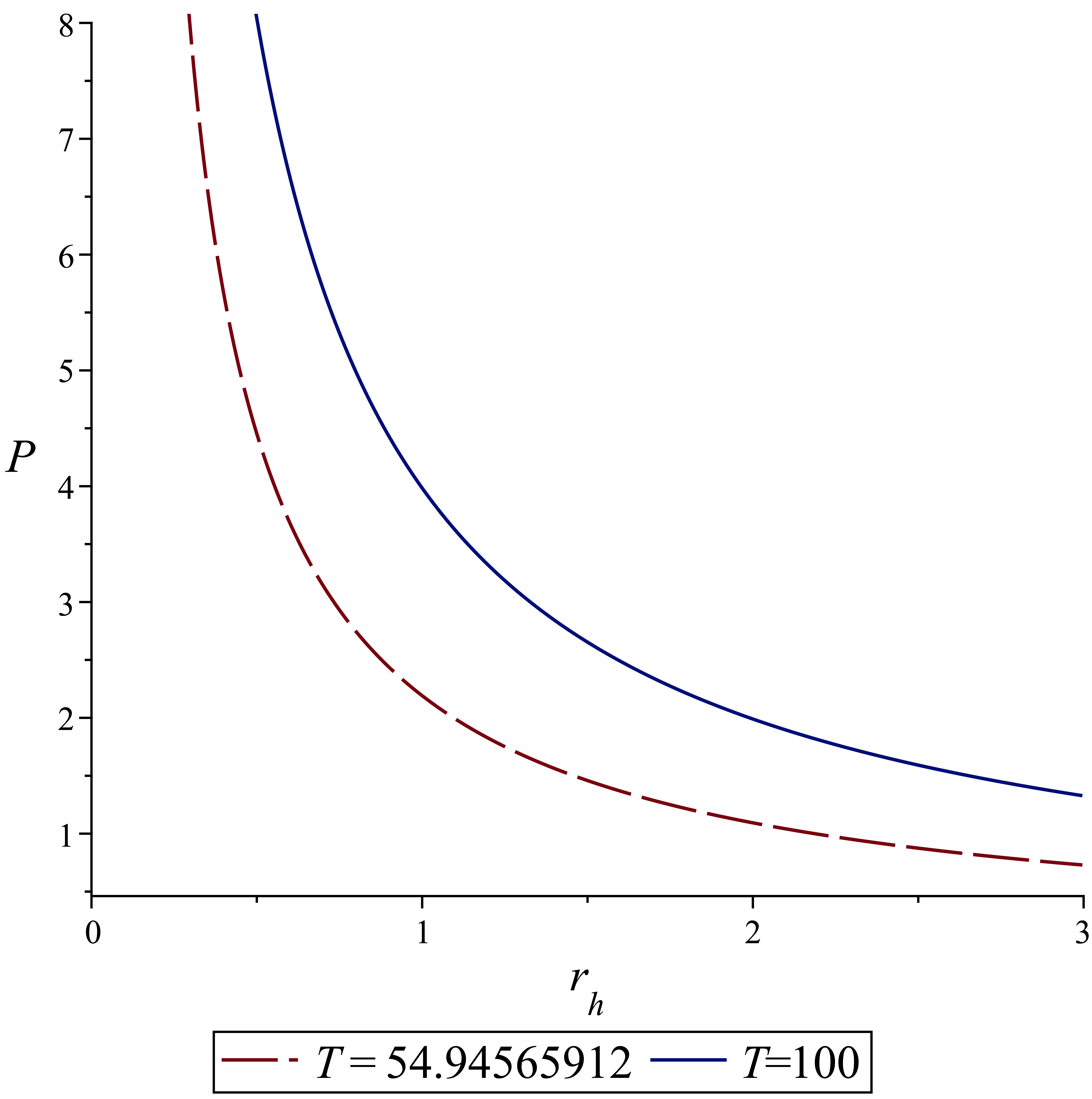}}
	\caption{$P-r_{h}$ diagrams for the values $ Q=0.1,q_1=-1$ and $q_2=0$ of the system. These diagrams show that for the uncoupled theory there is no critical behavior. Also, by increasing the temperature, the isotherms become more closer to the isotherms of an ideal gas.  \label{fig:0}}
\end{figure}

\begin{figure}[h!]
	\centering
	[a]{\includegraphics[width=6cm]{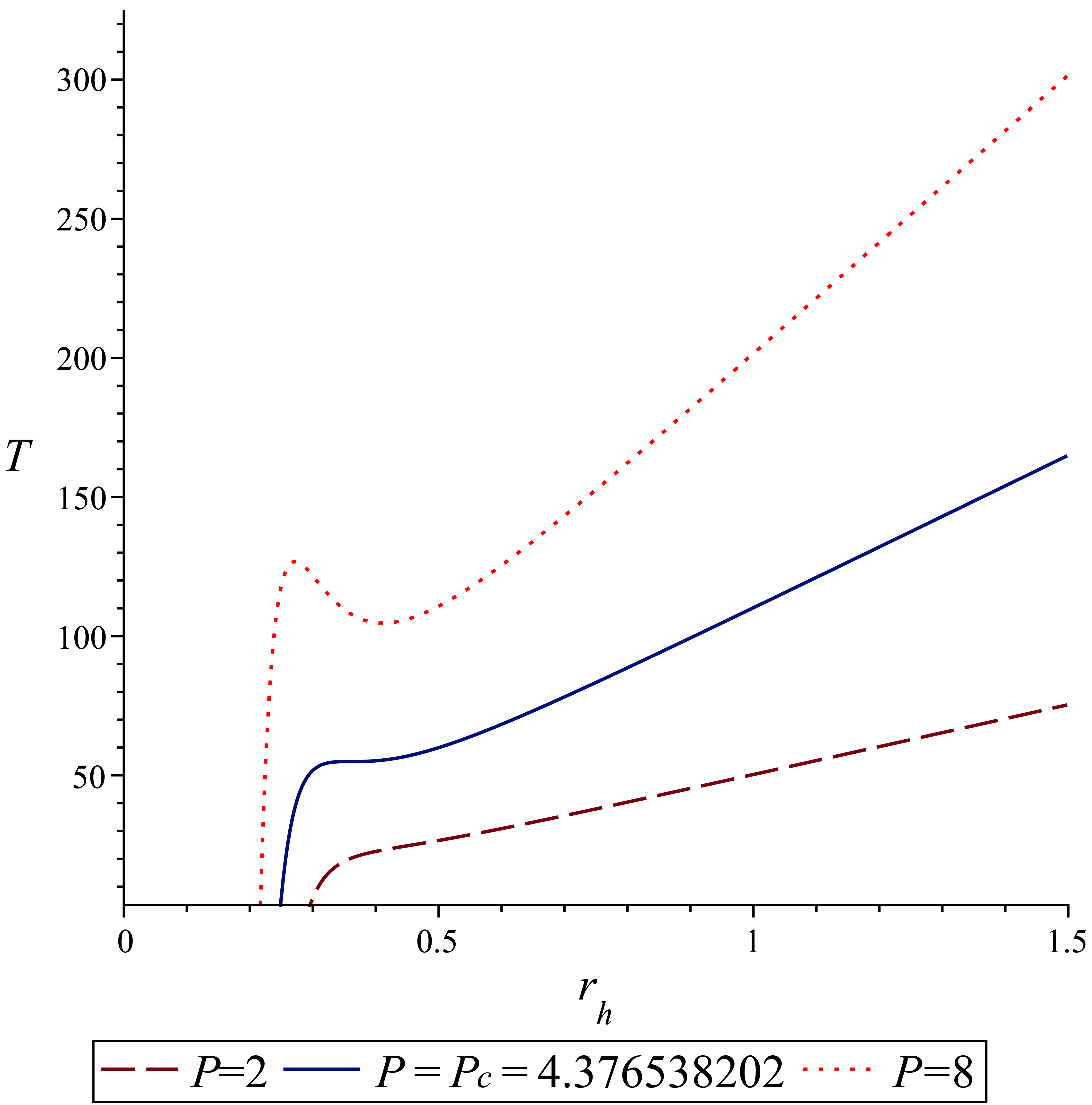}}\qquad
	[b]{\includegraphics[width=6cm]{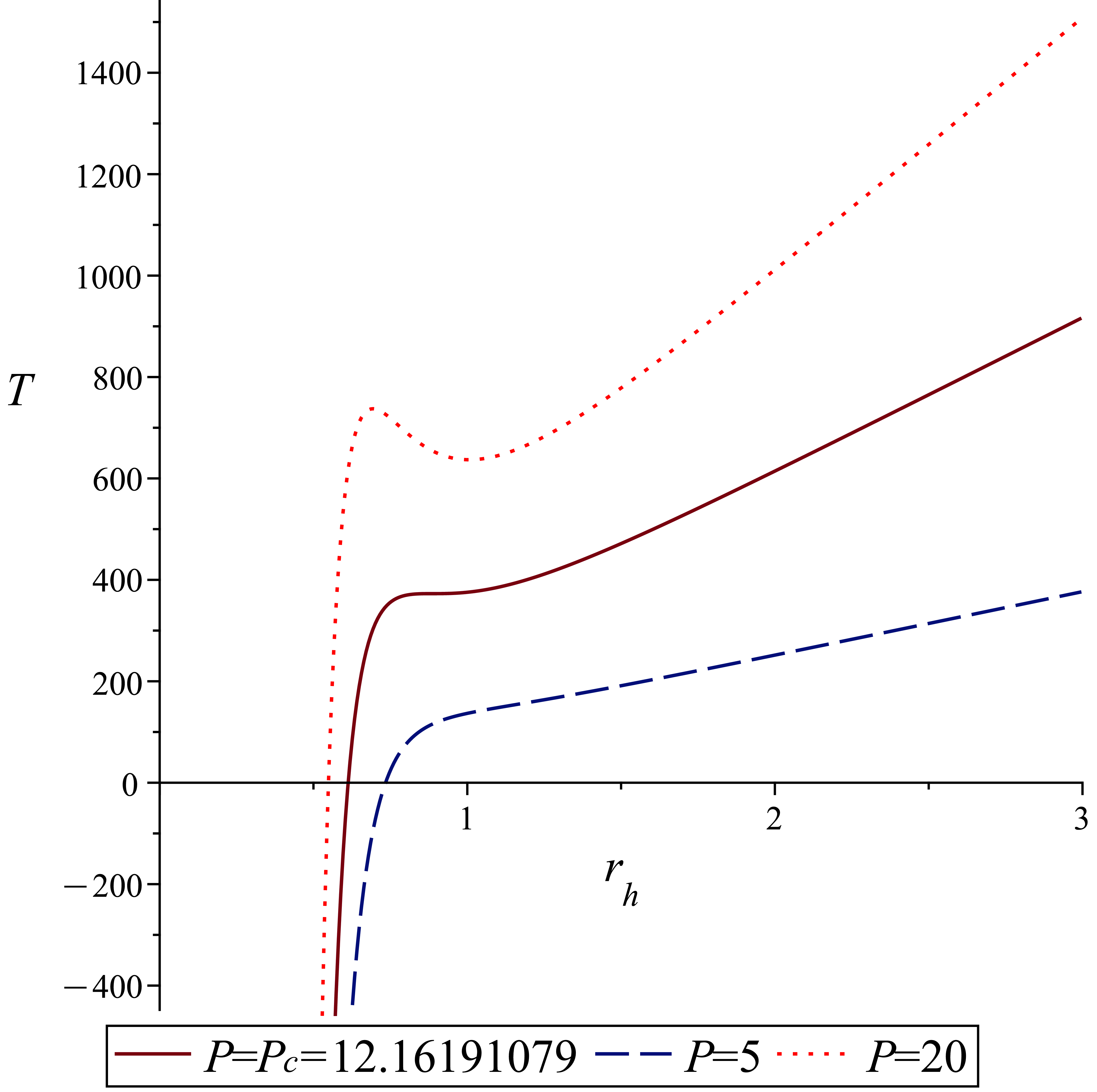}}
	\caption{$T-r_{h}$ diagrams of the system; (a):for $ Q=0.1,q_1=-1$ and $q_2=0.001$. The solid line represents the critical behavior of the system for the critical values $r_c=0.3598661919$ and $P_c=4.356738199$.,(b):for $Q=1,q_1=-1$ and $q_2=0.001$. The solid line represents the critical behavior of the system for the critical values $r_c=0.8868022599$ and $P_c=12.16191079$. \label{fig:4}}
\end{figure}

 
It can be easily seen that the uncoupled theory ($q_2=0$) has no critical behavior in this 4-dimensional model. See Fig (\ref{fig:0}). The pressure of the minimal theory is:
\begin{equation}\label{pun}
	P=-\frac{Q^{2} q_{1}}{2 r_{h}^{4}}+\frac{T}{8 \pi  r_{h}}.
\end{equation}
As the temperature of the system increases, the second term becomes dominant and the behavior of the system is closer to the behavior of an ideal gas.
The asymptotic relation for pressure is:
\begin{equation}\label{pasymp}
	P_{\text{asymp}}=\frac{T}{8 \pi  r_{h}}-\frac{Q^{2} q_{1}}{2 r_{h}^{4}}+3Q^{2} q_{2} \left(\frac{ T }{r_{h}^{5}}\right)+\mathrm{O}\! \left(\frac{1}{r_{h}^{6}}\right),
\end{equation}
The third term of this relation is the reason of unstable behavior of the system for high temperatures and small horizons. An ideal gas behavior is not achievable for ($T>T_c$), unless $q_2=0$. 
The diagrams of the temperature versus $r_h$, also show critical behaviors. See Fig. (\ref{fig:4}).
These diagrams show that for small horizons, the unstable behaviors appear even by increasing the temperature. This is due to the coupling parameter $q_2$.\par
\begin{figure}[h!]
	\centering
	[a]{\includegraphics[width=6cm]{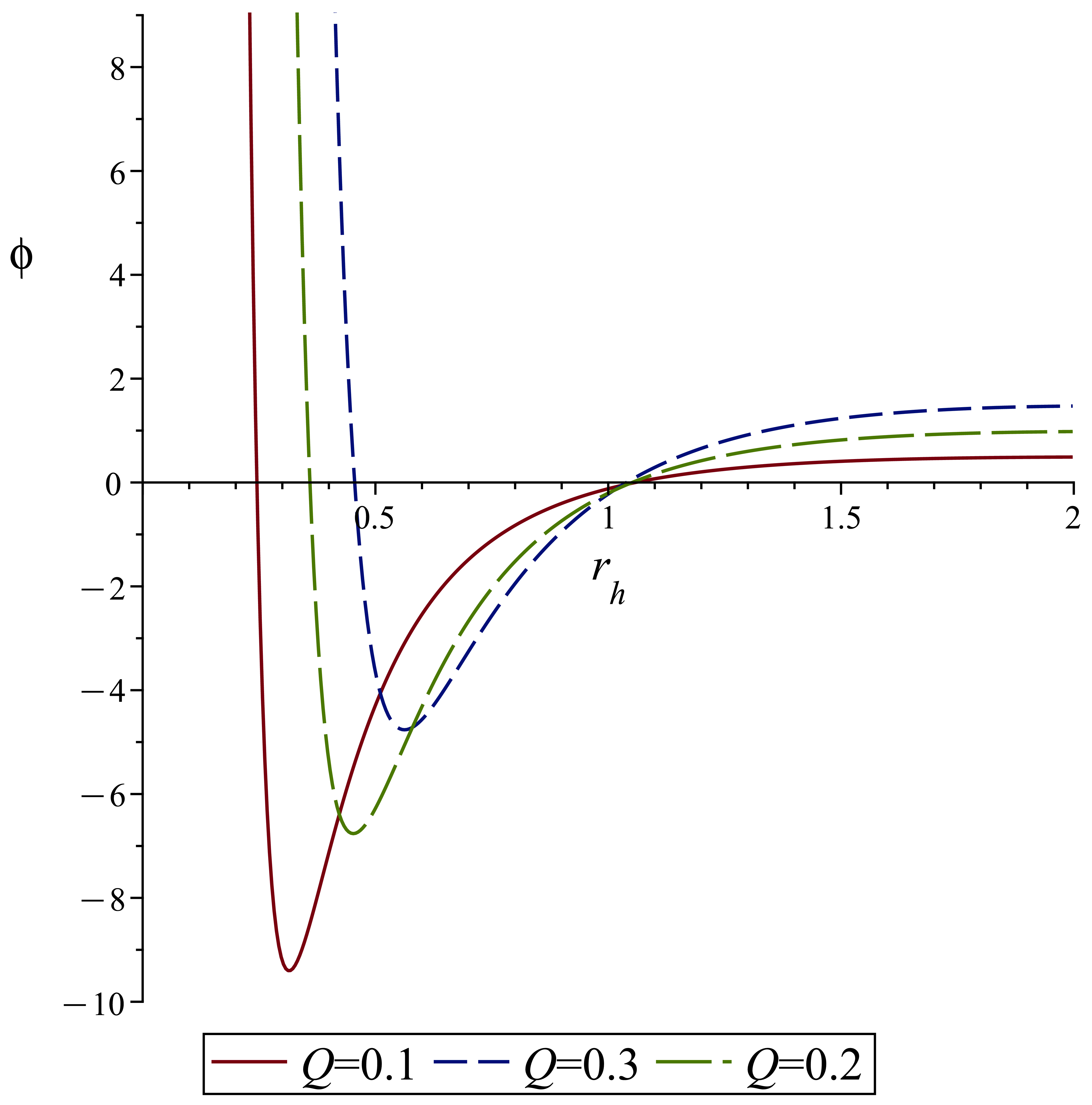}}\qquad
	[b]{\includegraphics[width=6cm]{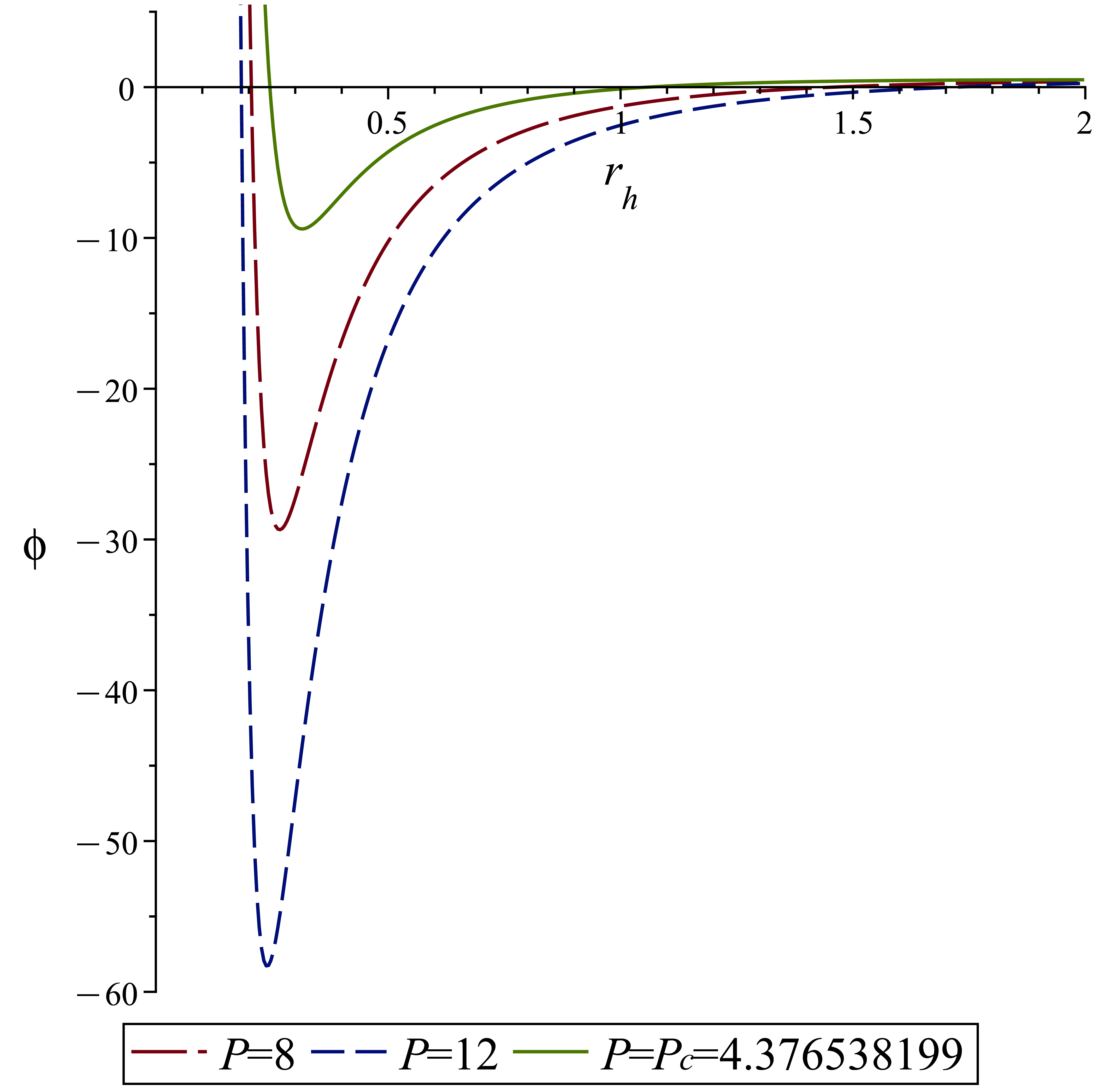}}
	\caption{$\phi-r_{h}$ diagram of the system; (a): for critical pressure $P=4.376538199$ and several values of the Yang-Mills charge,(b): for fixed Yang-Mills charge $Q=0.1$ and several values of the pressure. \label{fig:5}}
\end{figure}

We also depict the diagrams of the Yang-Mills potential in Fig(\ref{fig:5}) to have a better view of the Yang-Mills potential. In the left panel the pressure of the system is fixed while in the right panel the Yang-Mills charge is kept constant.
\par
\begin{figure}[h!]
	\centerline{\includegraphics[width=6cm]{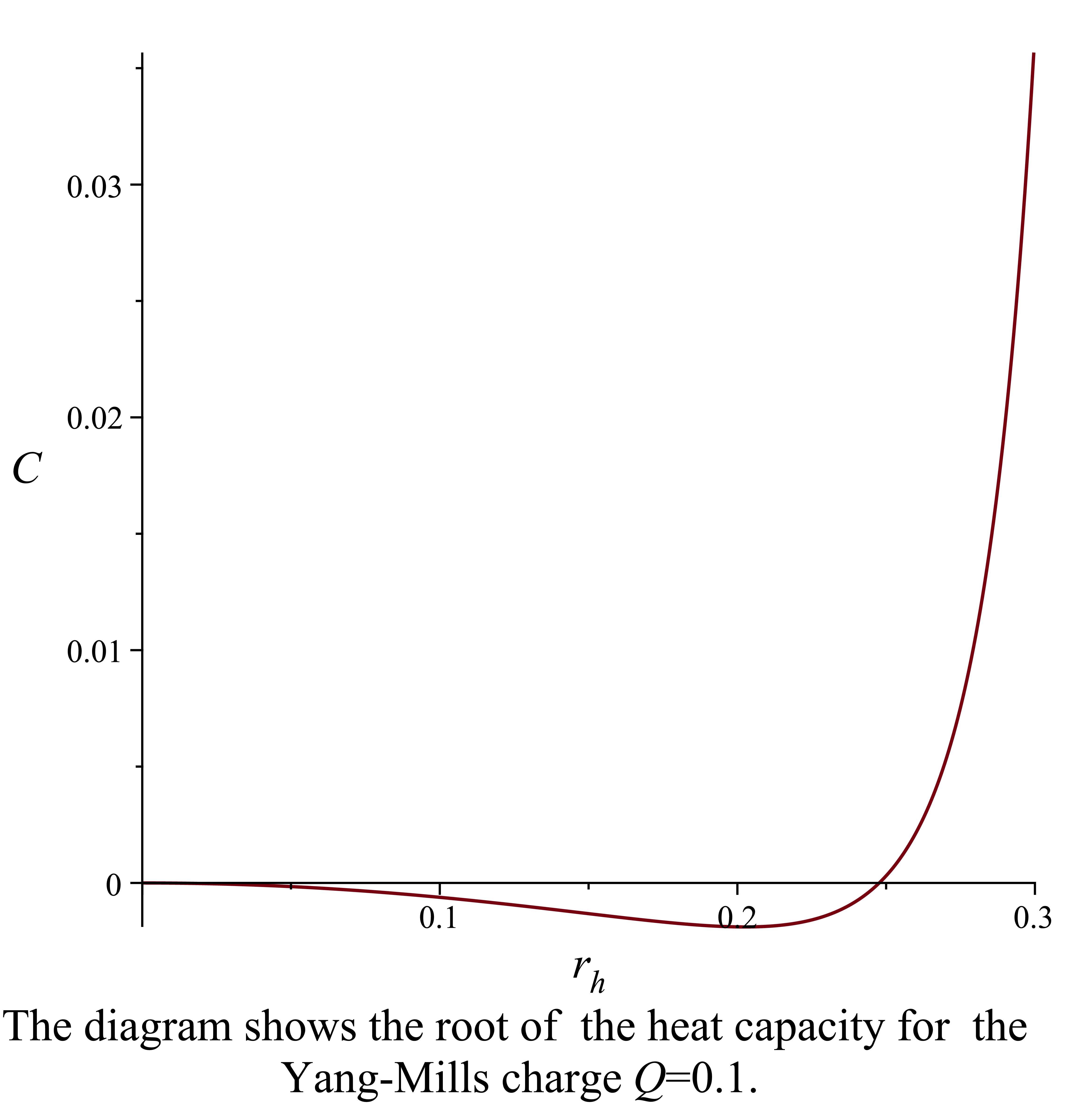}}
	\caption{This diagram shows the root of the heat capacity for $ Q=0.1$ and $0<r_h<0.3$ \label{fig:c0}}
\end{figure}
\begin{figure}[h!]
	\centering
	[a]{\includegraphics[width=6cm]{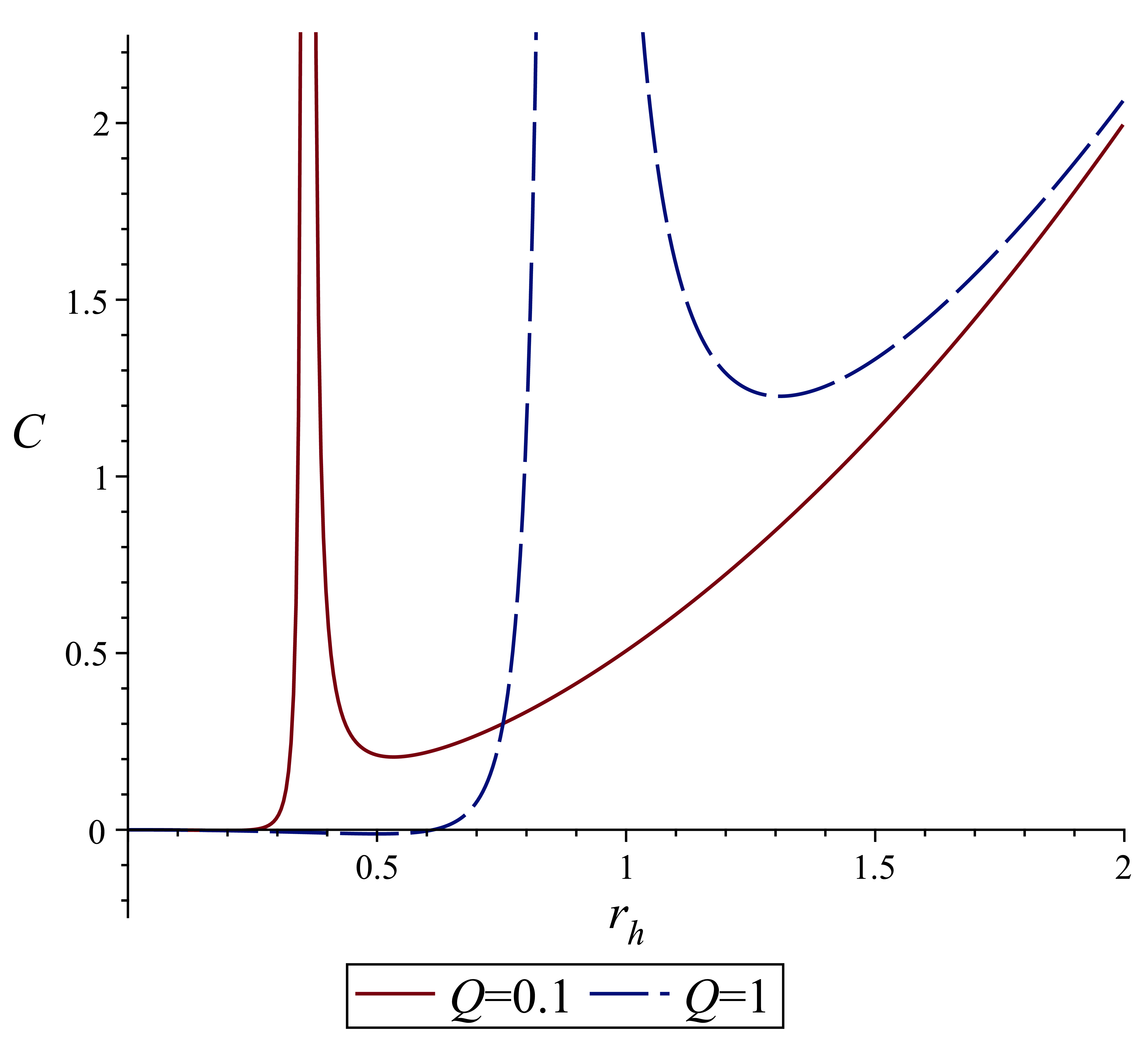}}\qquad
	[b]{\includegraphics[width=6cm]{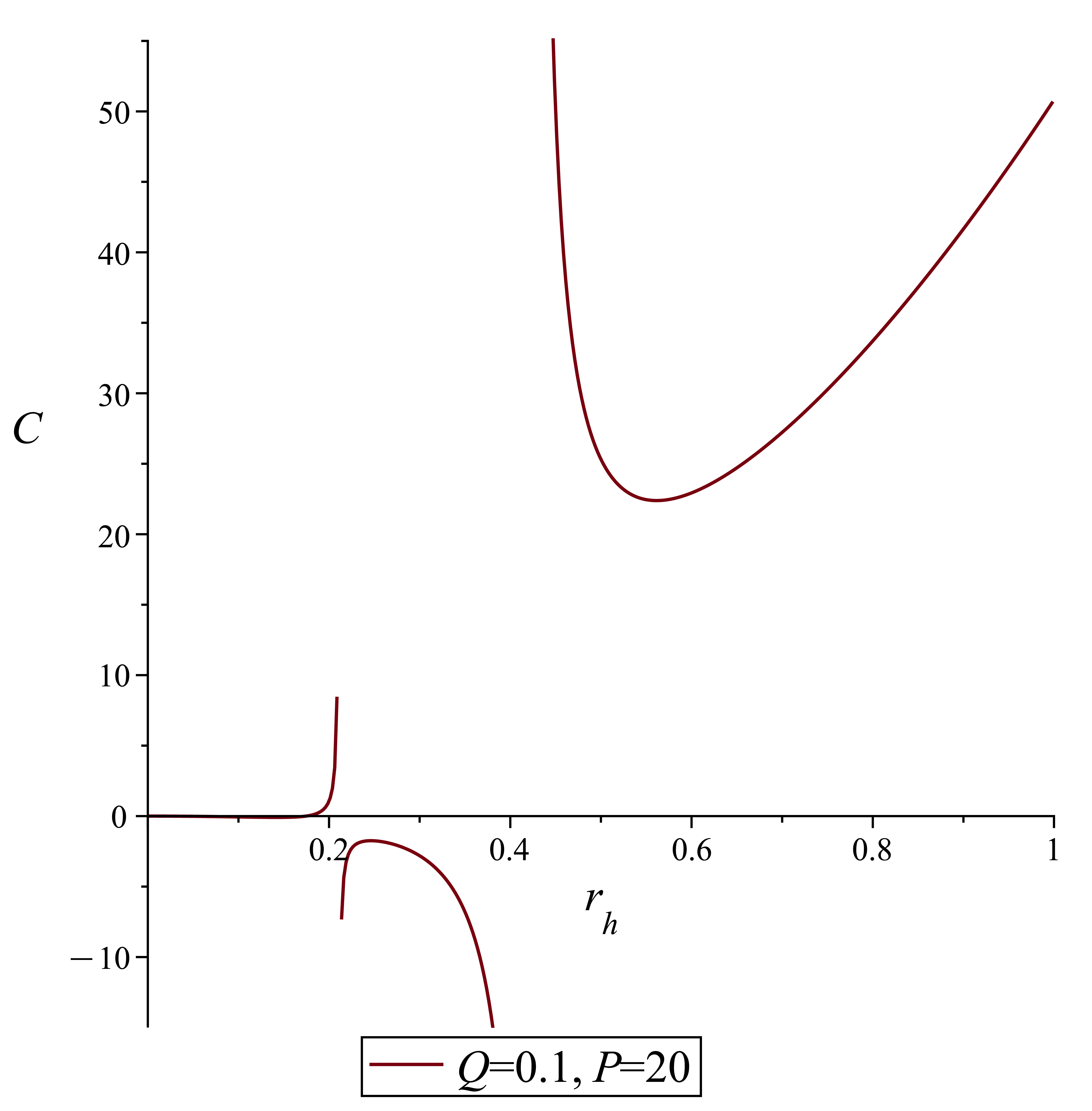}}
	\caption{$C-r_{h}$ diagrams of the system show unstable and stable regions of the system; (a):for $q_1=-1$, $q_2=0.001$ with critical pressures $P_c=4.356738199$ and $P_c=12.16191079$ and the Yang-Mills charges $Q=0.1$ and $Q=1$ respectively, (b): for pressure $P=20$ which is higher than the critical pressure and the Yang-Mills charge $Q=0.1$. \label{fig:6}}
\end{figure}

Now, we investigate the behavior of the heat capacity of the system at a fixed pressure i.e. $C_P$. As we know, the divergences and discontinuities in the heat capacity, imply on the existence of phase transitions. By using relation $C_P=\frac{T}{\left(\frac{\partial T}{\partial S}\right)_P}$, the heat capacity at constant pressure is obtained as follows:
\begin{equation}
	\begin{split}
		C_P&=\\
		&-\frac{r_{h}^{2} \left(\left(-384 P \pi^{2} Q^{2} r_{h}^{5}+2048 P \pi^{2} Q^{2} r_{h}^{4}-320 \pi^{2} Q^{4} q_{1} r_{h}+3840 \pi^{2} Q^{4} q_{1}\right) q_{2}+16 P \pi  r_{h}^{9}+8 \pi  Q^{2} q_{1} r_{h}^{5}\right)}{2 \left(-1152 P \pi^{2} Q^{2} r_{h}^{5}+8192 P \pi^{2} Q^{2} r_{h}^{4}-2240 \pi^{2} Q^{4} q_{1} r_{h}+30720 \pi^{2} Q^{4} q_{1}\right) q_{2}-32 P \pi  r_{h}^{9}+48 \pi  Q^{2} q_{1} r_{h}^{5}}.
	\end{split}
\end{equation}
Our investigations show that for $r_h< r_c$, the heat capacity has a root. But, the roots are in unphysical regions  for which the volume is negative. For example, the root of the heat capacity for the Yang-Mills charge $Q=0.1$, is $r_0=0.2476778090$. See Fig (\ref{fig:c0}). Remember that for $Q=0.1$, the volume is positive for $r_h>0.3498444712$. Thus, we can not discuss with certainty in this region. Between the root and $r_c$, the heat capacity is positive. But a divergence appears immediately. 
After the divergence point, the system transforms to the stable region for which $C_P>0$. See figure \ref{fig:6}. This shows that stability for $r_h<r_c$ strongly depends on the system parameters. The left panel of figure (\ref{fig:6}) shows that by increasing the Yang-Mills charge $Q$, the divergence points appear for larger horizons. The pressures of the curves in the left panel are the critical pressures. There are no intermediate regions in diagrams of heat capacity for critical pressures. For the second order phase transitions, the second derivative of the Gibbs function such as heat capacity are divergent at the critical points\cite{Huang,Callen}.
Thus the diagrams of the heat capacity for the critical pressures are signaling a second order phase transition.
Unfortunately, the investigation of critical exponents is almost impossible because of complicated dependence of quantities on $r_h$. It does not allow us to have analytical solutions or results to examine the associated functions to obtain the critical exponents.\par
\begin{figure}[h!]
	\centering
	[a]{\includegraphics[width=6cm]{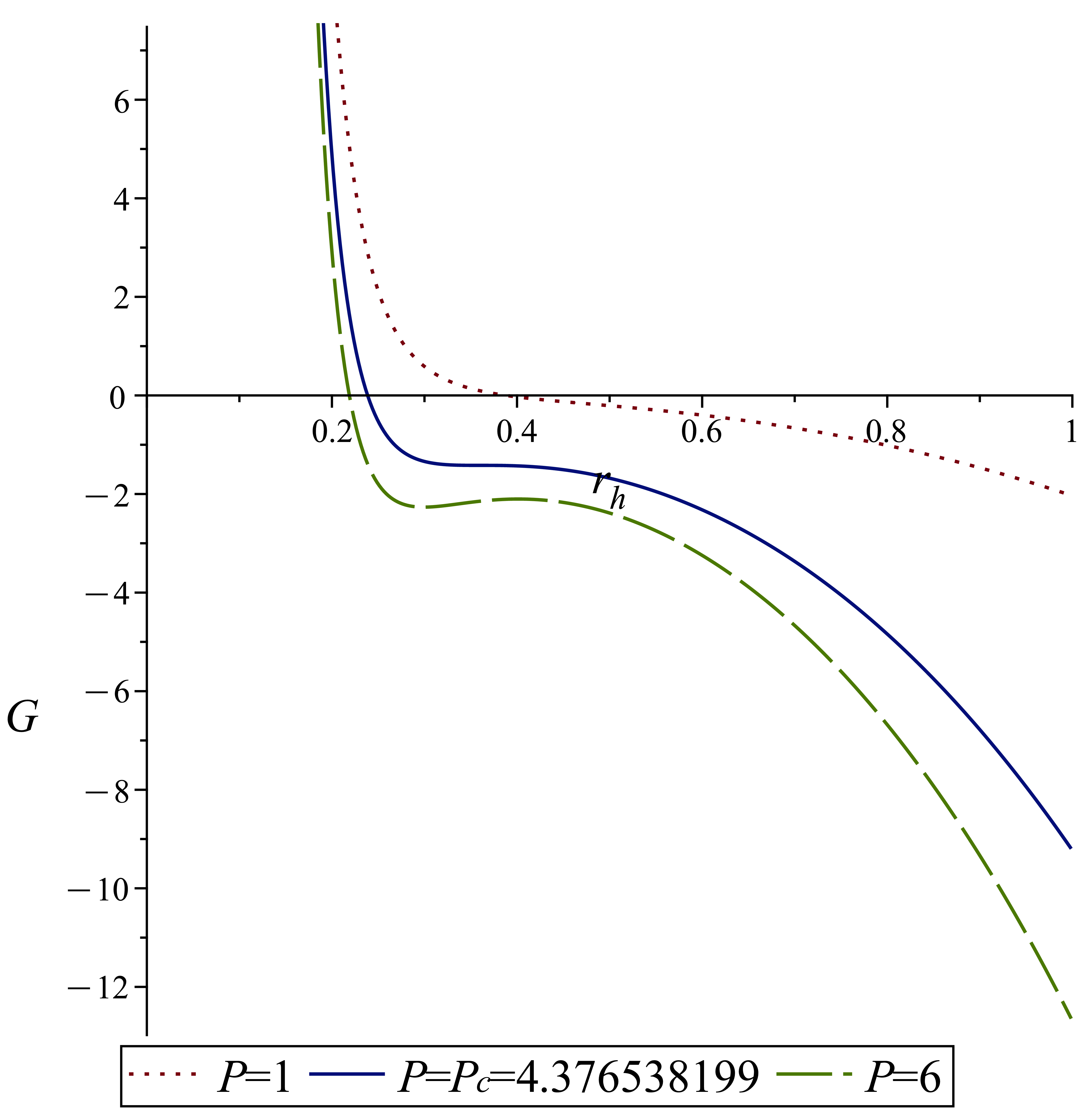}}\qquad
	[b]{\includegraphics[width=6cm]{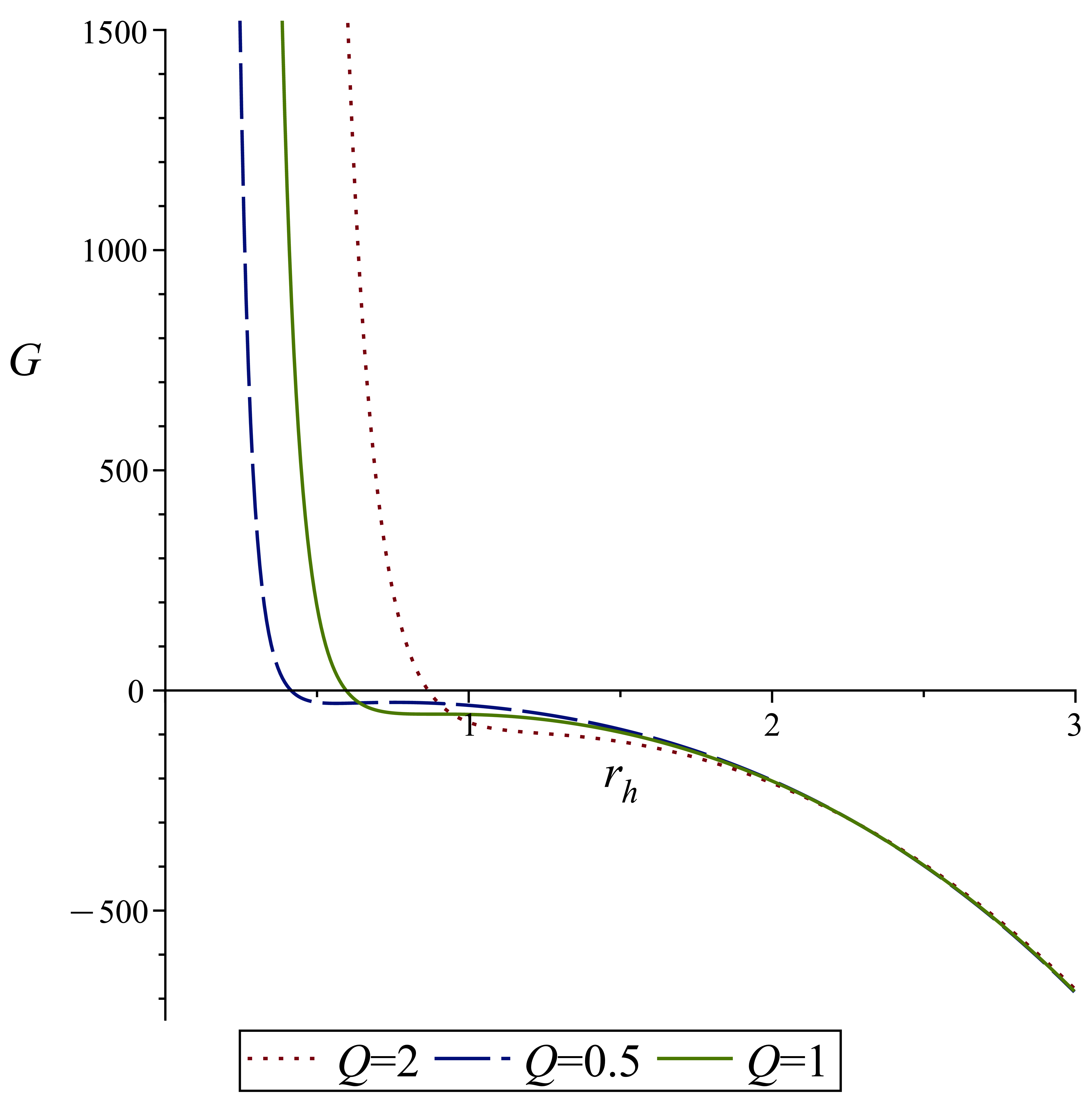}}
	\caption{$G-r_{h}$ diagrams of the system; (a):for $Q=0.1$  and several values of the pressure,(b):for the critical pressure $P_c=12.16191079$ and several values of the Yang-Mills charge $Q$. \label{fig:7}}
\end{figure}

 
For an arbitrary value $P=20$ and the Yang-Mills charge $Q=0.1$, the diagram of the heat capacity (right panel of figure (\ref{fig:7})), consists of three regions. Between the divergence points, $C_P<0$ and the system is unstable. While for the region after the second divergence point, $C_P>0$, and the system is stable.
Usually, to determine the order of transition of a thermodynamic system, the Ehrenfest criterion is used which states that for the first order phase transition the first derivatives (entropy and volume) of the Gibbs free energy is discontinuous. While, for the second order phase transition, the second derivative of the Gibbs function(proportional to the heat capacity, isothermal Compressibility and etc.) is divergent at critical points while the entropy and volume change continuously. The existence of a kink in the Gibbs function, which leads to a discontinuity in the first derivative of the Gibbs function with respect to the temperature(entropy), implies to the existence of the first order transition in this model. The Gibbs function in canonical ensemble($G=M-TS$), is as follows:
\begin{equation}\label{grp}
	\begin{split}
		&G(P,r_h)=\left(-\frac{640 \pi^{2} Q^{4} q_{1}}{r_{h}^{6}}+\frac{56 Q^{4} q_{1} \pi^{2}}{r_{h}^{5}}+\frac{144 Q^{2} P \pi^{2}}{r_{h}}-\frac{512 \pi^{2} Q^{2} P}{r_{h}^{2}}\right) q_{2}-\\
		&\frac{3 Q^{2} q_{1} \pi}{r_{h}}-\frac{2 P r_{h}^{3} \pi}{3}.
	\end{split}
\end{equation}
Figure (\ref{fig:7}) shows the plots of the Gibbs free energy versus $r_h$.
In the left panel the Yang-Mills charge is kept constant while in the right one the pressure is kept constant and is equal to the critical pressure $P_c=12.16191079$. The left panel shows that by increasing the pressure, the root of the Gibbs function occurs for smaller horizons which indicates that for higher pressures the system becomes more unstable.  The right panel, shows that for a fixed pressure, by increasing the value of Yang-Mills charge, the root of the Gibbs function occurs for larger horizons. For $G>0$, the system is unstable, while for the $G<0$, the system is stable globally. But, the new point in this study is the effect of coupling of the Yang-Mills field and gravity which makes the system unstable for small horizons even for temperatures higher than the critical temperatures. (Remember relation (\ref{pasymp}) and its third term.) The existence of $q_2$ is necessary to have phase transition. But, on the other hand it does not allow to have stable behavior above the critical point except for large horizons.\par
\begin{figure}[h!]
	\centerline{\includegraphics[width=7cm]{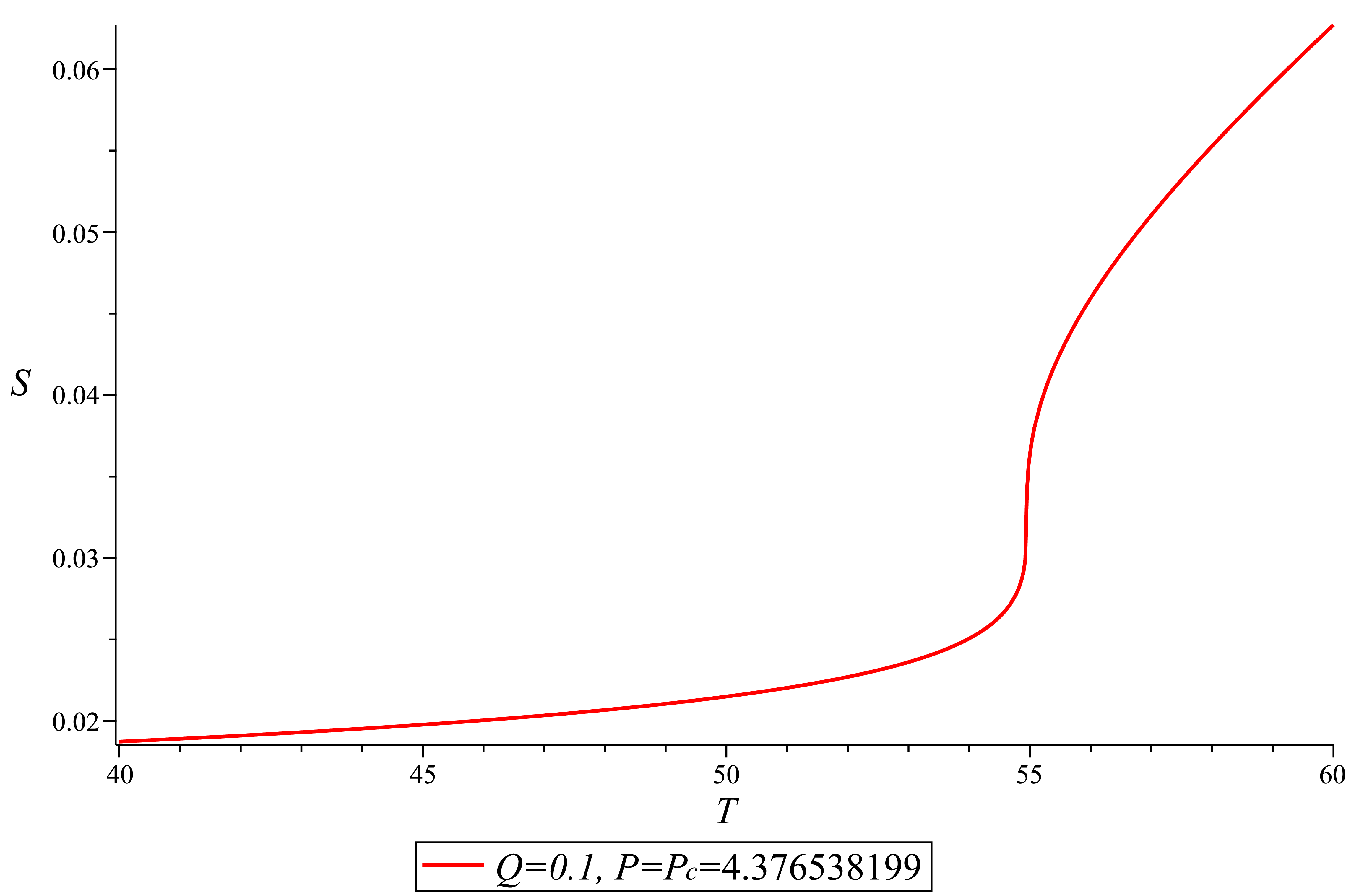}}
	\caption{$S-T$ diagram is continues at the critical temperature $T_c=54.94565912$. Thus, there is no first order phase transition in this system. \label{fig:8}}
\end{figure}

\begin{figure}[h!]
	\centering
	\includegraphics[width=6cm]{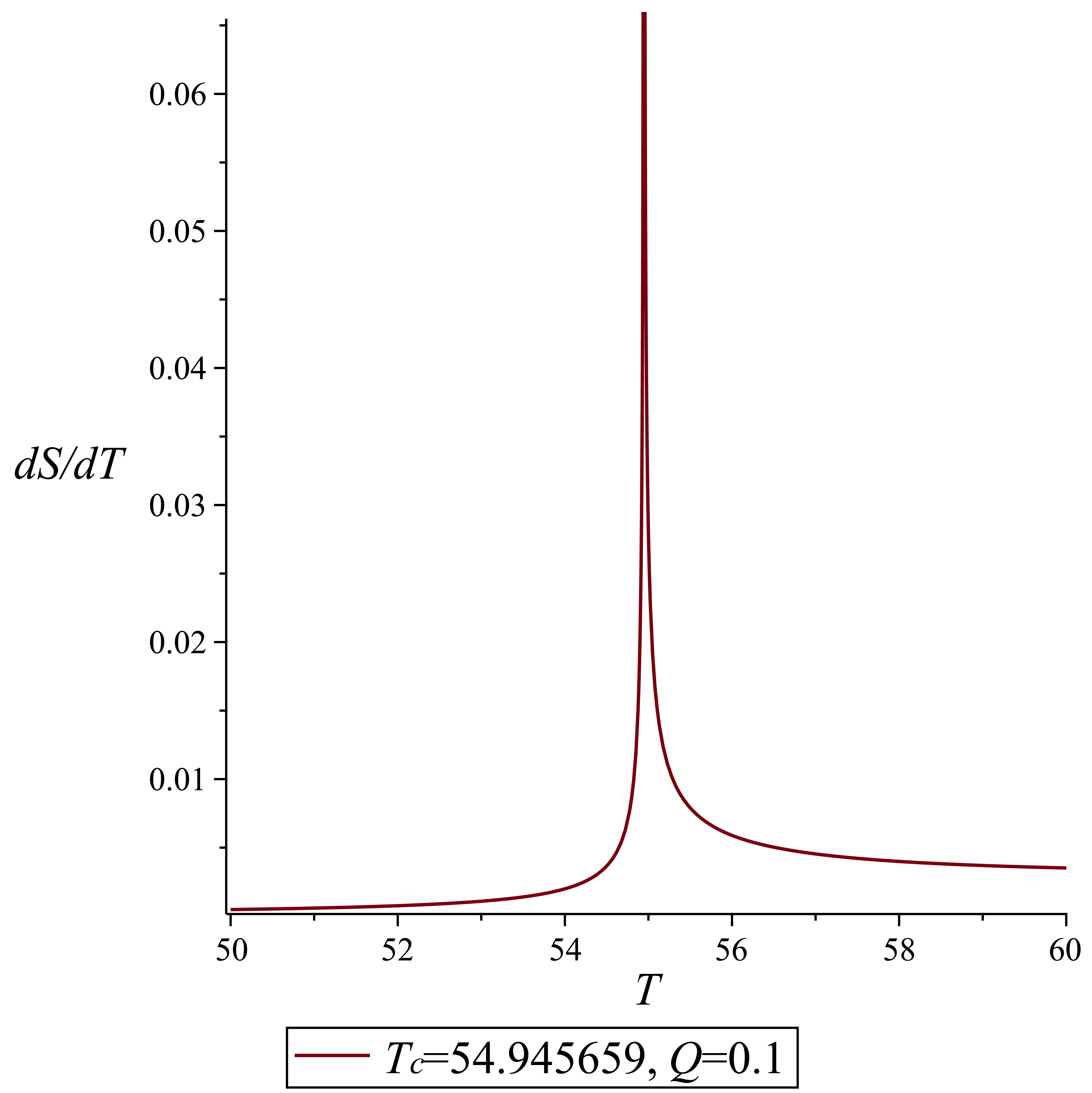}
	\caption{The diagram of $\frac{dS}{dT}$ versus $T$ shows a positive divergence at the critical temperature $T_c=54.94565912$. This signals a second order phase transition. \label{fig:9}}
\end{figure}

 
Before investigation the Gibbs function, we obtain the entropy diagram versus the temperature. Because, the entropy is the first derivative of the Gibbs function. Thus, any jump in the entropy diagram implies to existence of a kink in the Gibbs function and consequently the first-order phase transition.
The $S-T$ diagram for the critical temperature $T_c=54.94565912$ corresponding to the critical pressure $P_c=4.376538199$ is seen in Fig (\ref{fig:8}). The entropy is continuous and there is no jump or discontinuity at the critical temperature. So, we conclude that this first order phase transition is not seen in  this model.
The first derivative of the entropy with respect to the temperature which is proportional to the heat capacity must diverge at the critical point. The diagram of the first derivative of the entropy with respect to the temperature, shows a positive divergence which is a sign of the second order phase transition. See Fig. (\ref{fig:9}).
\begin{figure}[h!]
	\centering
	\includegraphics[width=6cm]{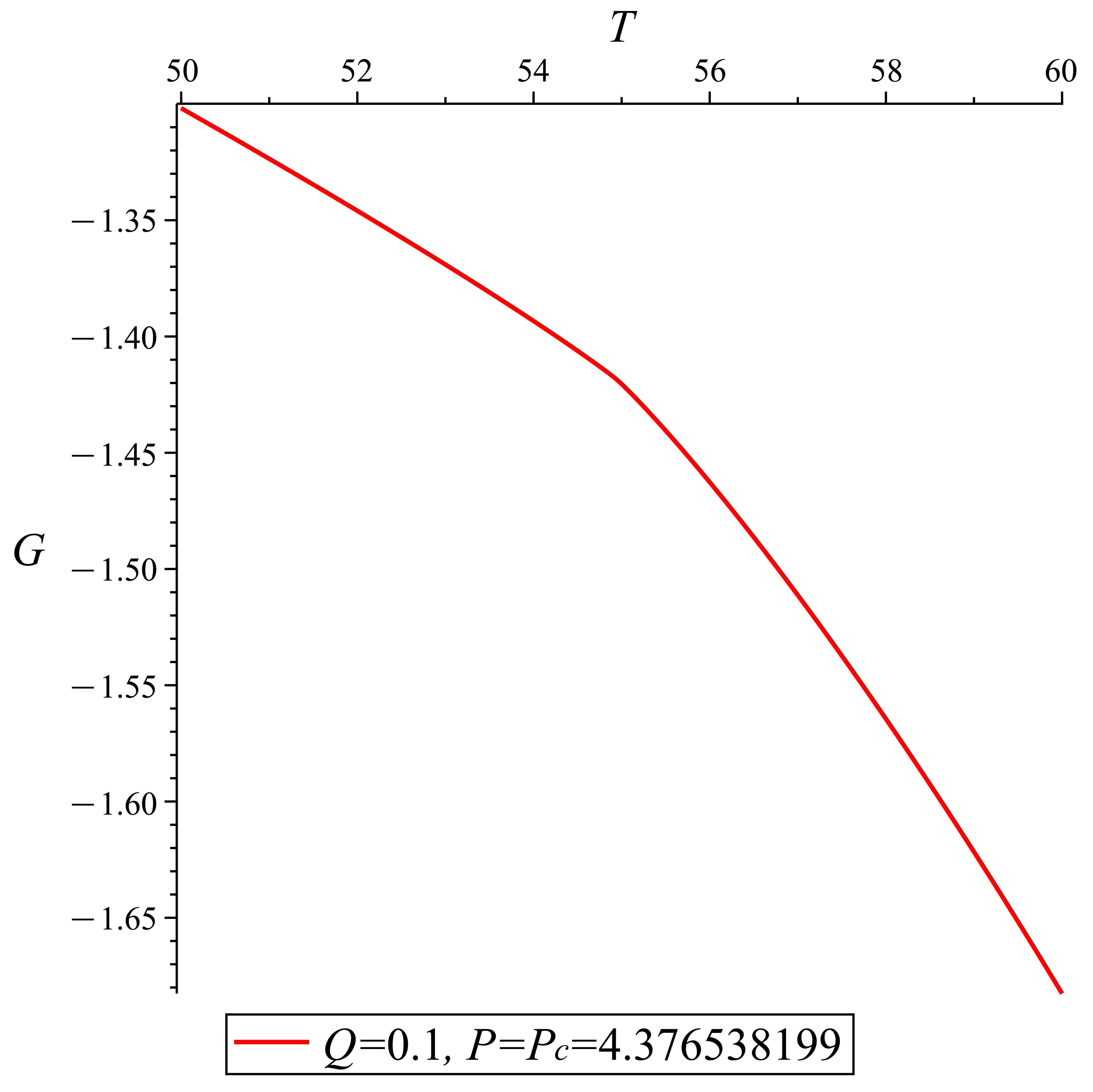}
	\caption{$G-T$ diagrams of the system at the critical pressure has a smooth behavior versus temperature. Thus the existence of a first order phase transition is ruled out. \label{fig:10}}
\end{figure}
A better zoom can be seen in the right panel of figure (\ref{fig:9}).
\par 
Now, we plot the the Gibbs function versus temperature to see its behavior at the critical temperature $T_c=54.94565912$. Since, the derivation an analytic relation for the Gibbs function in terms of temperature is impossible, we used a numeric investigation for the case $(Q=0.1, T_c=54.94565912, P_c=4.376538199)$.
Figure (\ref{fig:10}) shows the smooth behavior of the Gibbs energy versus temperature. Therefore, the possibility of a first order phase transition is ruled out.
This investigation shows that the first order phase transitions is not seen in this model(discontinuity of the entropy and existence of a kink in the Gibbs function). Also, at the critical point the heat capacity diagram diverges, indicating a second order phase transition.

\newpage
 \section{Conclusion}
In this paper, we introduced the non-minimal $RF^{(a)}_{\mu \alpha }F^{(a)\mu \alpha} $ black brane solution in AdS spacetime in four dimensions. Since, this model does not have general analytical solution, we solved it through a perturbation method in terms of non-minimal coupling coefficient $q_2$. 
The entropy of the system was obtained up to the first order of $q_2$ which brings a term proportional to the inverse of the horizon area. We ignored the inverse term by some estimations. Then, the mass or enthalpy of the system was obtained up to the first order of $q_2$. If we consider $q_2=0$, the phase transition will not be seen in this 4-dimensional model. By applying Ehrenfest's condition on the pressure of the system, we obtained inflection points for some arbitrary values of the Yang-Mills charge. However, deriving analytical results for the critical values of the thermodynamic quantities was not possible, we showed that the critical values are related to the Yang-Mills charge directly.  
Through the investigation of entropy we realized that the behavior of entropy versus temperature is smooth at critical point. It seems that the concavity of the entropy changes at the critical points which only signals a phase transition. On the other hand, the divergence of the heat capacity at critical point, specially a positive divergence, indicates a second order phase transition. Thus, this model suggests a phase transition from a normal phase to a superconducting phase in the dual field theory.\par 
This model has different behavior compared to the usual models in which the behavior of the system tends to the behavior of an ideal gas for temperatures higher than the critical temperature. For temperatures higher than the critical temperatures and large  horizons, the behavior of the system is close to the that of an ideal gas. But, for small horizons, the system becomes unstable even by increasing the temperature of the system. This is due to the coupling $q_2 RF^{(a)}_{\mu \alpha }F^{(a)\mu \alpha}$ and is interesting in its own way.

\vspace{1cm}
\noindent {\large {\bf Acknowledgment} } \\

We thank the respected referee who improved this paper with his/her useful comments. We also thank Dr. Manuel E. Rodrigues for introducing useful sources in this field.





\begin{thebibliography}{}

\bibitem{LIGOScientific:2021qlt}
R.~Abbott \textit{et al.} [LIGO Scientific, KAGRA and VIRGO],
``Observation of Gravitational Waves from Two Neutron Star\textendash{}Black Hole Coalescences,''
Astrophys. J. Lett. \textbf{915}, no.1, L5 (2021)
doi:10.3847/2041-8213/ac082e
[arXiv:2106.15163 [astro-ph.HE]].


\bibitem{Sadeghi:2021qou}
M.~Sadeghi,
``Non-abelian Einstein-Born-Infeld AdS black brane and color DC conductivity,''
doi:10.1007/s12648-022-02317-z
[arXiv:2111.12916 [hep-th]].


\bibitem{Sadeghi:2022mog}
M.~Sadeghi,
``AdS black brane coupled to non-abelian logarithmic gauge theory and color DC conductivity,''
doi:10.1139/cjp-2023-0150
[arXiv:2203.05023 [hep-th]].

\bibitem{Balakin:2005fu}
A.~B.~Balakin and J.~P.~S.~Lemos,
``Non-minimal coupling for the gravitational and electromagnetic fields: A General system of equations,''
Class. Quant. Grav. \textbf{22}, 1867-1880 (2005)
doi:10.1088/0264-9381/22/9/024
[arXiv:gr-qc/0503076 [gr-qc]].




\bibitem{Balakin:2015gpq}
A.~B.~Balakin, J.~P.~S.~Lemos and A.~E.~Zayats,
``Regular nonminimal magnetic black holes in spacetimes with a cosmological constant,''
Phys. Rev. D \textbf{93}, no.2, 024008 (2016)
doi:10.1103/PhysRevD.93.024008
[arXiv:1512.02653 [gr-qc]].





\bibitem{Bamba:2008ja}
K.~Bamba and S.~D.~Odintsov,
``Inflation and late-time cosmic acceleration in non-minimal Maxwell-$F(R)$ gravity and the generation of large-scale magnetic fields,''
JCAP \textbf{04}, 024 (2008)
doi:10.1088/1475-7516/2008/04/024
[arXiv:0801.0954 [astro-ph]].



\bibitem{Hawking:1982dh}
S.~W.~Hawking and D.~N.~Page,
``Thermodynamics of Black Holes in anti-De Sitter Space,''
Commun. Math. Phys. \textbf{87}, 577 (1983)
doi:10.1007/BF01208266.


\bibitem{Davies:1978zz}
P.~C.~W.~Davies,
``Thermodynamics of black holes,''
Rept. Prog. Phys. \textbf{41}, 1313-1355 (1978).

\bibitem{Davies:1989ey}
P.~C.~W.~Davies,
``Thermodynamic Phase Transitions of {Kerr-Newman} Black Holes in De Sitter Space,''
Class. Quant. Grav. \textbf{6}, 1909 (1989).

\bibitem{Yan:2021uzw}
D.~W.~Yan, Z.~R.~Huang and N.~Li,
``Hawking-Page phase transitions of charged AdS black holes surrounded by quintessence,''
Chin. Phys. C \textbf{45}, no.1, 015104 (2021)
doi:10.1088/1674-1137/abc0cf


\bibitem{Ghanaatian:2019xhi}
M.~Ghanaatian, M.~Sadeghi, H.~Ranjbari and G.~Forozani,
``Effects of the external string cloud on the Van der Waals like behavior and efficiency of AdS-Schwarzschild black holes in massive gravity,''
Mod. Phys. Lett. A \textbf{35}, no.24, 2050203 (2020)
doi:10.1142/S021773232050203X
[arXiv:1906.00369 [hep-th]].


\bibitem{Sadeghi:2023tzf}
M.~Sadeghi and F.~Rahmani,
``The phase transition of Rastall AdS black hole with cloud of strings and quintessence,''
Int. J. Mod. Phys. A \textbf{38}, no.20, 2350102 (2023)
doi:10.1142/S0217751X23501026
[arXiv:2301.12411 [hep-th]].


\bibitem{Wu:2020fij}
B.~Wu, C.~Wang, Z.~M.~Xu and W.~L.~Yang,
``Ruppeiner geometry and thermodynamic phase transition of the black hole in massive gravity,''
Eur. Phys. J. C \textbf{81}, no.7, 626 (2021)
doi:10.1140/epjc/s10052-021-09407-y
[arXiv:2006.09021 [gr-qc]]


\bibitem{Chabab:2019mlu}
M.~Chabab, H.~El Moumni, S.~Iraoui and K.~Masmar,
``Phase transitions and geothermodynamics of black holes in dRGT massive gravity,''
Eur. Phys. J. C \textbf{79}, no.4, 342 (2019)
doi:10.1140/epjc/s10052-019-6850-0
[arXiv:1904.03532 [hep-th]].

\bibitem{Kumar:2023gjt}
A.~Kumar, A.~Sood, J.~K.~Singh, A.~Beesham and S.~G.~Ghosh,
``Phase structure and critical behaviour of charged-AdS black holes with perfect fluid dark matter,''
Phys. Dark Univ. \textbf{40}, 101220 (2023)
doi:10.1016/j.dark.2023.101220

\bibitem{KordZangeneh:2016btt}
M.~Kord Zangeneh, A.~Dehyadegari, M.~R.~Mehdizadeh, B.~Wang and A.~Sheykhi,
``Thermodynamics, phase transitions and Ruppeiner geometry for Einstein\textendash{}dilaton\textendash{}Lifshitz black holes in the presence of Maxwell and Born\textendash{}Infeld electrodynamics,''
Eur. Phys. J. C \textbf{77}, no.6, 423 (2017)
doi:10.1140/epjc/s10052-017-4989-0
[arXiv:1610.06352 [hep-th]].


\bibitem{Mahapatra:2016dae}
S.~Mahapatra,
``Thermodynamics, Phase Transition and Quasinormal modes with Weyl corrections,''
JHEP \textbf{04}, 142 (2016)
doi:10.1007/JHEP04(2016)142
[arXiv:1602.03007 [hep-th]].

\bibitem{Ranjbari:2019ktp}
H.~Ranjbari, M.~Sadeghi, M.~Ghanaatian and G.~Forozani,
``Critical behavior of AdS Gauss\textendash{}Bonnet massive black holes in the presence of external string cloud,''
Eur. Phys. J. C \textbf{80}, no.1, 17 (2020)
doi:10.1140/epjc/s10052-019-7592-8
[arXiv:1911.10803 [hep-th]].





\bibitem{Hendi:2017mfu}
S.~H.~Hendi and M.~Momennia,
``Reentrant phase transition of Born\textendash{}Infeld-dilaton black holes,''
Eur. Phys. J. C \textbf{78}, no.10, 800 (2018)
doi:10.1140/epjc/s10052-018-6278-y
[arXiv:1709.09039 [gr-qc]].


\bibitem{Chamblin:1999tk}
A.~Chamblin, R.~Emparan, C.~V.~Johnson and R.~C.~Myers,
``Charged AdS black holes and catastrophic holography,''
Phys. Rev. D \textbf{60}, 064018 (1999)
doi:10.1103/PhysRevD.60.064018
[arXiv:hep-th/9902170 [hep-th]].


\bibitem{Chamblin:1999hg}
A.~Chamblin, R.~Emparan, C.~V.~Johnson and R.~C.~Myers,
``Holography, thermodynamics and fluctuations of charged AdS black holes,''
Phys. Rev. D \textbf{60}, 104026 (1999)
doi:10.1103/PhysRevD.60.104026
[arXiv:hep-th/9904197 [hep-th]].

\bibitem{Wei:2010yw}
S.~W.~Wei, Y.~X.~Liu, Y.~Q.~Wang and H.~Guo,
``Thermodynamic Geometry of Black Hole in the Deformed Horava-Lifshitz Gravity,''
EPL \textbf{99}, no.2, 20004 (2012)
doi:10.1209/0295-5075/99/20004
[arXiv:1002.1550 [hep-th]].


\bibitem{Maldacena}
J. M. Maldacena, 
``The Large N limit of superconformal field theories and supergravity,''
Int.\ J.\ Theor.\ Phys.\  {\bf 38} (1999) 1113 [Adv.\ Theor.\ Math.\ Phys.\  {\bf 2} (1998) 231] [hep-th/9711200].



\bibitem{Aharony}
O. Aharony, S.~S.~Gubser, J.~M.~Maldacena, H.~Ooguri and Y.~Oz,
``Large N field theories, string theory and gravity,''
Phys.\ Rept.\  {\bf 323}, 183 (2000)
[hep-th/9905111].



\bibitem{Witten:1998qj}
E.~Witten,
``Anti-de Sitter space and holography,''
Adv. Theor. Math. Phys. \textbf{2}, 253-291 (1998)
doi:10.4310/ATMP.1998.v2.n2.a2
[arXiv:hep-th/9802150 [hep-th]].

\bibitem{Witten:1998zw}
E.~Witten,
``Anti-de Sitter space, thermal phase transition, and confinement in gauge theories,''
Adv. Theor. Math. Phys. \textbf{2}, 505-532 (1998)
doi:10.4310/ATMP.1998.v2.n3.a3
[arXiv:hep-th/9803131 [hep-th]].




\bibitem{Gubser:2005ih}
Gubser S.~S.~,
``Phase transitions near black hole horizons,''
\textit{Class. Quant. Grav.} \textbf{22} 5121(2005).

\bibitem{Gubser:2008px}
Gubser S.~S.~,
``Breaking an Abelian gauge symmetry near a black hole horizon,''
\textit{Phys. Rev. D} \textbf{78} 065034 (2008).

\bibitem{Shepherd:2015dse} 
B.~L.~Shepherd and E.~Winstanley,
``Dyons and dyonic black holes in ${\mathfrak {su}}(N)$ Einstein-Yang-Mills theory in anti-de Sitter spacetime,''
Phys.\ Rev.\ D {\bf 93}, no. 6, 064064 (2016)
doi:10.1103/PhysRevD.93.064064
[arXiv:1512.03010 [gr-qc]].

\bibitem{Sadeghi:2023hxd}
M.~Sadeghi,
``Holographic aspects of non-minimal RF\ensuremath{\mu}\ensuremath{\alpha}(a)F(a)\ensuremath{\mu}\ensuremath{\alpha} black brane,''
Mod. Phys. Lett. A \textbf{38}, no.20n21, 2350098 (2023)
doi:10.1142/S0217732323500980
[arXiv:2302.07247 [hep-th]].

\bibitem{Deh1}
M.K.~ Zangeneh, A.~ Sheykhi, M.H.~ Dehghani, Thermodynamics of charged rotating dilaton black branes with power-law Maxwell field. Eur. Phys. J. C \text{75}, 497 (2015).

\bibitem{Huang}
K.~ Huang, Statistical Mechanics, John Wily and Sons (1987).


\bibitem{Callen}
H.B~ Callen, Thermodynamics and an introduction to thermostatistics, John Wily and Sons (1985).



  
  
\end{thebibliography}
\end{document}